\documentclass{elsart}

\usepackage{cite,graphicx,amsmath,amsfonts,amssymb,mathrsfs}

\newcommand{\ie}{{\it i.e.}}

\newcommand{\eg}{{\it e.g.}}

\newcommand{\eq}{Eq.}
\newcommand{\eqs}{Eqs.}
\newcommand{\fig}{Fig.}
\newcommand{\figs}{Figs.}
\newcommand{\Ref}{Ref.}
\newcommand{\Refs}{Refs.}
\newcommand{\Sec}{Sec.~}
\newcommand{\Secs}{Secs.~}
\newcommand{\App}{App.~}

\begin{document}

\begin{frontmatter}

\begin{flushright}
{\small TUM-HEP-467/02}
\end{flushright}

\title{A Flavor Symmetry Model for Bilarge Leptonic Mixing and the
  Lepton Masses}

\author[labela]{Tommy Ohlsson\thanksref{label0}},
\thanks[label0]{E-mail: tohlsson@ph.tum.de}
\author[labela]{Gerhart Seidl\thanksref{label1}}
\thanks[label1]{E-mail: gseidl@ph.tum.de}

\address[labela]{Institut f{\"u}r Theoretische Physik, Physik-Department,
Technische Universit{\"a}t M{\"u}nchen, James-Franck-Stra{\ss}e, 85748
Garching bei M{\"u}nchen, Germany}

\date{\today}
         
\begin{abstract}
We present a model for leptonic mixing and the lepton masses
based on flavor symmetries and higher-dimensional mass operators. The
model predicts bilarge leptonic mixing (\ie, the mixing angles
$\theta_{12}$ and $\theta_{23}$ are large and the mixing angle
$\theta_{13}$ is small) and an inverted hierarchical neutrino mass
spectrum. Furthermore, it approximately yields the experimental
hierarchical mass spectrum of the charged leptons. The obtained values
for the leptonic mixing parameters and the neutrino mass squared
differences are all in agreement with atmospheric neutrino data, the
Mikheyev--Smirnov--Wolfenstein large mixing angle solution of the
solar neutrino problem, and consistent with the upper bound on the
reactor mixing angle. Thus, we have a large, but not close to
maximal, solar mixing angle $\theta_{12}$, a nearly maximal
atmospheric mixing angle $\theta_{23}$, and a small reactor mixing
angle $\theta_{13}$. In addition, the model predicts $\theta_{12}
\simeq \tfrac{\pi}{4} - \theta_{13}$.
\end{abstract}

\end{frontmatter}

\newpage

\section{Introduction}

The fermionic sector of the standard model (SM) of elementary particle
physics is described by 13 renormalized parameters (6 quark
masses, 3 charged lepton masses, 3 CKM mixing angles\footnote{The
  mixing angles in the quark sector are usually called the
  Cabibbo--Kobayashi--Maskawa (CKM) mixing angles
  \cite{Cabibbo:1963yz,Kobayashi:1973fv}.}, and one CP violation
phase). Obviously, these
parameters are not arbitrary, but exhibit relations which can only be
understood when going beyond the SM. One possibility to obtain
realistic quark masses and CKM mixing angles in an extension of the SM is
to introduce flavor symmetries that are sequentially broken. At the
first glance, the hierarchical
mass spectra of the quarks and the charged leptons actually suggest underlying
non-Abelian flavor symmetry groups acting on the first and second
generations only.\footnote{By placing the first two generations into
  irreducible representations of flavor symmetries, one can in
  supersymmetric models achieve near degeneracy of the corresponding
  squark masses thus suppressing large flavor changing neutral
  currents \cite{Dermisek:1999vy,King:2001uz}.}
However, in the light of recent atmospheric
\cite{Fukuda:1998mi,Fukuda:1998ah,Fukuda:2000np,Toshito:2001dk,Shiozawa:2002} and
solar
\cite{Fukuda:2001nj,Smy:2001wf,Fukuda:2002pe,Smy:2002,Ahmad:2001an,Ahmad:2002jz,Ahmad:2002ka,Hallin:2002}
neutrino experimental results, it seems to be difficult
to extend this idea to the neutrinos.
Especially, the result that, among the different possible solutions of
the solar neutrino problem, the Mikheyev--Smirnov--Wolfenstein (MSW)
\cite{mikh85,mikh86,wolf78}
large mixing angle (LMA) solution is the presently preferred one
\cite{Bahcall:2001zu,Bahcall:2001cb,Bahcall:2002hv}\footnote{Actually,
  there exist several recent global solar neutrino
  oscillation analyses including the latest SNO data that strongly
  favor the MSW LMA solution of the solar neutrino problem, see, \eg,
  \Refs\cite{Ahmad:2002ka,Barger:2002iv,Bandyopadhyay:2002xj,Bahcall:2002hv,Aliani:2002ma,deHolanda:2002pp}. However, we have chosen to list and use
  the values obtained in \Ref\cite{Bahcall:2002hv}.},
draws a picture of the involved flavor symmetries and their
breaking mechanisms that differs remarkably from the early approaches,
which have been applied to the quark sector. In the ``standard''
parameterization, the MSW LMA solution implies that we have a {\it bilarge}
mixing scenario in the lepton sector in which the solar mixing angle
$\theta_{12}$ is large, but not necessarily close to maximal, the
atmospheric mixing angle $\theta_{23}$ is nearly maximal, and the
reactor mixing angle $\theta_{13}$ is small. Clearly, this is in
sharp contrast to the quark sector in which all mixing angles are
small \cite{groo00} and it indicates that the flavor symmetries act on the
third generation too.

By assuming only an Abelian U(1) flavor symmetry, one obtains that the
atmospheric mixing angle may be large, but cannot be enforced to be
nearly maximal 
\cite{Ibanez:1994ig,Binetruy:1995ru,Shafi:2000su}. Therefore, a
natural close to maximal $\nu_\mu$-$\nu_\tau$-mixing can be
interpreted as a strong hint for some underlying non-Abelian flavor
symmetry acting on the second and third generations
\cite{Mohapatra:1998ka,Wetterich:1998vh}.
Neutrino mass models which give large or maximal solar and atmospheric mixing
angles by putting the second and third generations of the leptons
into the regular representation of the symmetric group ${\rm S_2}$
\cite{Grimus:2001ex} or into the irreducible two-dimensional
representation of the 
symmetric group ${\rm S_3}$ \cite{Mohapatra:1999zr} are, in general,
plagued with a fine-tuning problem in the charged lepton sector, since
they tend to predict the muon and tau masses to be of the same order
of magnitude, \ie, they lack providing an understanding of the
hierarchical mass spectrum in the charged lepton sector. A recently
proposed model based on an SU(3) flavor symmetry \cite{King:2001uz}
gives approximately bimaximal leptonic mixing as well as a successful
description of the charged fermion masses, but predicts the presently
disfavored MSW low mass (LOW) or vacuum oscillation (VAC) solution of
the solar neutrino problem. Similarly, the highly predictive models of
flavor democracy
\cite{Fritzsch:1996dj,Fritzsch:1998xs,Fritzsch:1999ee,Fukugita:1998vn,Fukugita:1998kt,Tanimoto:1998yz,Kang:1998gs,Tanimoto:1999pj}
yield large solar and atmospheric mixing angles, but fit the LOW or
VAC solution rather than the MSW LMA solution \cite{Dorsner:2001sg}.
In grand unified theory model building, it seems that
the MSW LMA solution with a normal hierarchical neutrino mass spectrum
is more natural than with an inverted one \cite{King:2002} (for a recent
phenomenological analysis of minimal schemes for the MSW LMA solution
with inverted hierarchical neutrino mass spectra, see, \eg,
\Ref\cite{He:2002rv}). Also from empirical lepton and quark mass
spectra analyses a normal hierarchical (or inverse hierarchical) neutrino mass
spectrum seems to be rather plausible \cite{Lindner:2001kd}. A
comparably simple way of generating the MSW LMA solution with normal
hierarchical neutrino mass spectra is, \eg, provided by models based
on single right-handed neutrino dominance \cite{King:2002nf}.

\begin{figure}[ht!]
\begin{center}
\includegraphics*[height=10cm]{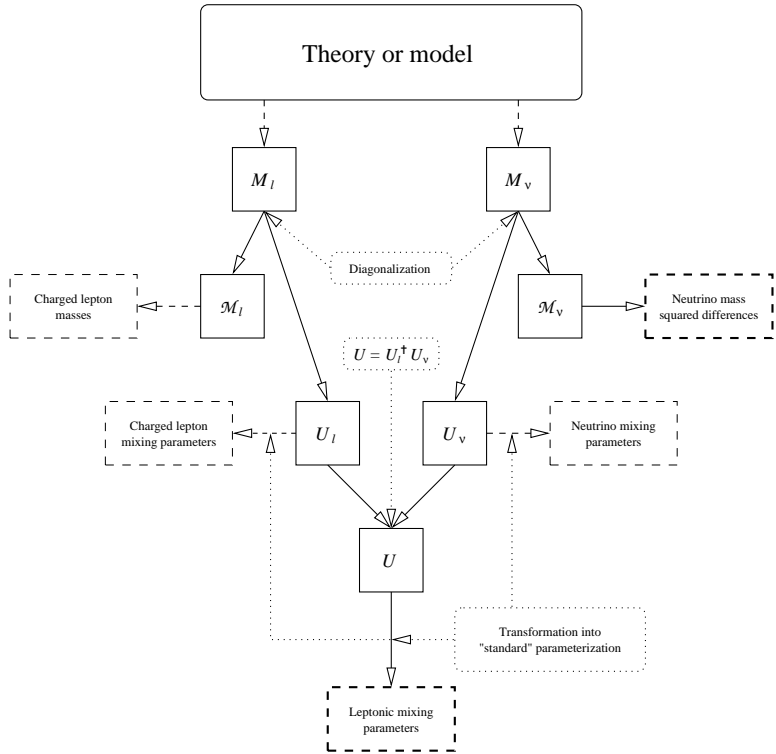}
\end{center}
\caption{Flow-scheme for computing leptonic mixing parameters and
  lepton masses from any given charged lepton and neutrino mass matrices.}
\label{fig:scheme}
\end{figure}

In a previous Letter \cite{Ohlsson:2002na}, we introduced a model for
bilarge leptonic mixing based on higher-dimensional operators, using
the Froggatt--Nielsen mechanism, and Abelian horizontal flavor
symmetries of continuous and discrete types.
In this paper, we consider a modified and extended version of this model
and we explicitly demonstrate the {\it vacuum alignment mechanism},
which produces a nearly maximal atmospheric mixing angle
$\theta_{23}$ as well as a large, but not close to maximal, solar
mixing angle $\theta_{12}$, as required by the MSW LMA
solution. Simultaneously, the vacuum alignment mechanism generates a strictly
hierarchical charged
lepton mass spectrum thus resolving the fine-tuning problem many realistic
models, which seek to predict the MSW LMA solution, are suffering from.
Furthermore, this model gives a small mixing of the first and second
generations of the charged leptons, which is comparable with the
mixing of the quarks, whereas the mixing among the neutrinos is essentially
bimaximal \cite{barg98}. The actual leptonic mixing angles are then a result of
combining the contributions coming from both the charged leptons and
the neutrinos (see \fig~\ref{fig:scheme}). Thus, the model predicts
the relation $\theta_{12} \simeq \tfrac{\pi}{4} - \theta_{13}$ between
the solar mixing angle $\theta_{12}$ and the reactor mixing angle
$\theta_{13}$, which is non-zero and lies in the range of the quark mixing
angles.

Note that our study assumes that there are three neutrino flavors, and
therefore, three neutrino flavor states $\nu_\alpha$ ($\alpha =
e,\mu,\tau$) and also three neutrino mass eigenstates $\nu_a$ ($a =
1,2,3$). Furthermore, it assumes that all $\mathcal{CP}$ violation phases are
equal to zero.

The paper is organized as follows: In \Sec\ref{sec:rep_con}, we
introduce the representation content of our model including U(1)
charges and discrete symmetries. The multi-scalar potential of the
model is then analyzed in \Sec\ref{sec:scalar_pot}, where we explicitly
demonstrate the vacuum alignment mechanism. Next, in
\Secs\ref{sec:Yukawa_cl} and \ref{sec:Yukawa_n}, the Yukawa
interactions of the charged leptons and neutrinos, respectively, are
investigated and discussed, which lead to the mass matrices of the
corresponding particles. In \Sec\ref{sec:massmix}, the lepton mass matrices are
diagonalized yielding the charged lepton masses and the neutrino mass
squared differences as well as the charged lepton and neutrino mixing
angles (see again \fig~\ref{fig:scheme}). In \Sec\ref{sec:leptonic},
the total leptonic mixing angles are derived and
calculated. Implications for neutrinoless double $\beta$-decay,
astrophysics, and cosmology are briefly studied in \Sec\ref{sec:impl}.
Finally, in \Sec\ref{sec:s&c}, we present a summary as well as our
conclusions.
In addition, we present in the Appendix a scheme for how to transform any
given $3 \times 3$ unitary matrix to the ``standard'' parameterization
form of the Particle Data Group \cite{groo00}.

\section{The representation content}
\label{sec:rep_con}

Let us consider an extension of the SM in which the lepton masses are
generated by higher-dimensional operators \cite{wein79,wilc792} via
the Froggatt--Nielsen mechanism \cite{frog79}. (A classification of
effective neutrino mass operators has been given in
\Ref\cite{Babu:2001ex}.) Since we are mainly concerned with the
question of whether there is a possible naturally maximal
$\nu_\mu$-$\nu_\tau$-mixing in the MSW LMA solution or not, for which
the properties of the quarks are seemingly irrelevant, we will, for
simplicity and without loss of generality, omit the quark sector in
our further discussion. For a recent related study including also the
quark sector, see, \eg, \Refs\cite{Froggatt:2002tb,Nielsen:2002cw}.
In a self-explanatory notation, we will denote the lepton doublets as
$L_\alpha = ( \begin{matrix} \nu_{\alpha L} & e_{\alpha L}
\end{matrix} )$, where $\alpha = e,\mu,\tau$, and the
right-handed charged leptons as $E_\alpha = e_{\alpha R}$, where
$\alpha = e,\mu,\tau$. The part of the scalar sector, which carries non-zero 
SM quantum numbers, consists of two Higgs doublets $H_1$ and $H_2$,
where $H_1$ couples to the neutrinos and $H_2$ to the charged
leptons. This can be achieved by assuming, \eg, a discrete
${\mathbb{Z}}_2$ symmetry under which $H_2$ and $E_\alpha$ ($\alpha =
e,\mu,\tau$) are odd and $H_1$ and the rest of the SM fields are
even. The masses of the charged leptons arise through the mixing with
additional heavy right-handed charged fermions, which all have masses
of the order of magnitude of some characteristic mass scale
$M_1$. Apart from some general prescriptions of their transformations
under the flavor symmetries, which will be introduced below, it is not
necessary to explicitly present the fundamental theory of these
additional (or extra) charged fermions.
This is in contrast to the neutrino sector, which we will extend by five
additional heavy SM singlet Dirac neutrinos $N_e$, $N_\mu$, $N_\tau$,
$F_1$, and $F_2$. The neutrinos $F_1$ and $F_2$ are supposed to have
masses of the same order $M_1$ as the charged intermediate
Froggatt--Nielsen states, whereas $N_e$, $N_\mu$, and $N_\tau$ all
have masses of the order of magnitude of some relevant high (unification) mass
scale $M_2$. While $M_2$ takes the role of some seesaw scale
\cite{gell79,yana79,Mohapatra:1980ia}
(and it is therefore responsible for the smallness of the neutrino
masses), $M_1$ can be as low as several TeV
\cite{Leurer:1993wg,Perez-Lorenzana:2001pr}. In order to
obtain the structures of the lepton mass matrices from an underlying
symmetry principle in the context of a renormalizable field theory, we
will furthermore extend the scalar sector by additional SM singlet
scalar fields $\phi_1, \phi_2, \ldots, \phi_{10}$, $\phi'_1, \phi'_2, \ldots,
\phi'_6$, and $\theta$ and we will assign these fields gauged
horizontal U(1) charges $Q_1$, $Q_2$, and $Q_3$ as follows:
$$
{\tiny
\begin{tabular}{c|c}
 & $(Q_1,Q_2,Q_3)$\\
\hline
$L_e$,$E_e$ & $(1,0,0)$\\
$L_\mu$,$L_\tau$,$E_\mu$,$E_\tau$ & $(0,1,0)$\\
$N_e$ & $(1,0,0)$\\
$N_\mu,N_\tau$ & $(0,1,0)$\\
$F_1$ & $(1,0,0)$\\
$F_2$ & $(-1,0,1)$\\
\hline
$H_1,H_2$       & $(0,0,0)$\\
$\phi_1,\phi_2$ & $(1,-1,2)$\\
$\phi_3,\phi_4$ & $(0,0,0)$\\
$\phi_5,\phi_6$ & $(0,0,1)$\\
$\phi_1',\phi_2',\phi_3',\phi_4',\phi_5',\phi_6'$ & $(0,0,0)$\\
$\phi_7,\phi_8$ & $(-1,-1,0)$\\
$\phi_9$          & $(-2,0,1)$\\
$\phi_{10}$       & $(0,0,0)$\\
$\theta$          & $(0,0,-1)$
\end{tabular}
}
$$
In the rest of the paper, it will always be understood that the Higgs
doublets $H_1$ and $H_2$ are total singlet under transformations of other
additional symmetries. Note that the charges $Q_1$ and $Q_2$ are
anomalous. However, it is known that anomalous U(1) charges may arise
in effective field theories from strings. Then, the cancellation of the
anomalies must be accomplished by the Green--Schwarz mechanism
\cite{Green:1984sg}.

The first generation of the charged leptons is distinguished from the
second and third generations if we require
for $P\equiv {\rm e}^{2\pi{\rm i}/n}$, where the integer $n$ obeys
$n\geq 5$, invariance of the Lagrangian under transformation of the
following ${\mathbb{Z}}_n$ symmetry:
\begin{eqnarray}
{\mathcal D}_1 & : & \left\{
\begin{matrix}
E_e\rightarrow P^{-4} E_e,& E_{\mu} \rightarrow P^{-1} E_{\mu},&
E_{\tau} \rightarrow P^{-1} E_{\tau},\\
\phi_1 \rightarrow P\phi_1, & \phi_2 \rightarrow
 P\phi_2,\\ \phi_3 \rightarrow P\phi_3, & 
\phi_4 \rightarrow P\phi_4,\\
\phi_5\rightarrow P\phi_5, & \phi_6\rightarrow P\phi_6,
\end{matrix}\right.
\end{eqnarray}
where we assume that the fields $N_e$, $N_\mu$, $N_\tau$, $F_1$, and
$F_2$ are singlets under transformation of the symmetry
${\mathcal{D}}_1$. In addition, the symmetry ${\mathcal{D}}_1$ forbids
the fields $\phi_1, \phi_2, \ldots, \phi_6$ to participate in the
leading order mass terms for the neutrinos. Furthermore, the permutation
symmetries
\begin{eqnarray}
{\mathcal D}_2 & : & \left\{
\begin{matrix}
L_\mu \rightarrow - L_\mu, & E_\mu \rightarrow -E_\mu,\\
N_\mu \rightarrow -N_\mu,&\\
\phi_1'\leftrightarrow \phi_2', & \phi_1 \leftrightarrow \phi_2,\\
\phi_7\rightarrow -\phi_7,&
\end{matrix}\right.\\
\nonumber\\
{\mathcal D}_3 & : & \left\{
\begin{matrix}
L_\mu \rightarrow - L_\mu, & N_\mu \rightarrow -N_\mu,\\
\phi_3'\leftrightarrow \phi_4', & \phi_3\leftrightarrow\phi_4,\\
\phi_5'\leftrightarrow \phi_6', & \phi_5\leftrightarrow\phi_6,\\
\phi_7\rightarrow -\phi_7,
\end{matrix}\right.\\
\nonumber\\
{\mathcal D}_4 & : & \left\{
\begin{matrix}
L_\mu \leftrightarrow L_\tau, & E_\mu \leftrightarrow E_\tau,\\
N_\mu\leftrightarrow N_\tau, &\\
\phi_2 \rightarrow -\phi_2, & \phi_4 \rightarrow -\phi_4, &
\phi_6\rightarrow -\phi_6,\\ \phi_7\leftrightarrow \phi_8
\end{matrix}\right. 
\end{eqnarray}
are responsible for generating a naturally maximal atmospheric mixing
angle, since they establish exact degeneracies of the Yukawa couplings
in the leptonic 2-3-subsector. These permutation symmetries also play
a crucial role in the scalar sector in which they restrict some of the
couplings in the multi-scalar potential to be exactly degenerate (at
tree level), which means that degenerate vacuum expectation values
(VEVs) can emerge after spontaneous symmetry breaking (SSB). This so-called
vacuum alignment mechanism can work if we assume the
discrete symmetries
\begin{eqnarray}
{\mathcal D}_5 & : &
\left\{ 
\begin{matrix}
\phi_1' \rightarrow -\phi_1', & \phi_1 \rightarrow -\phi_1,\\
\phi_3' \rightarrow -\phi_3', & \phi_3 \rightarrow -\phi_3,\\
\phi_5' \rightarrow -\phi_5', & \phi_5 \rightarrow -\phi_5,
\end{matrix}\right.\\
\nonumber\\
{\mathcal D}_6& : & 
\left\{
\begin{matrix}
 E_e\rightarrow P^{-(4l+1)}E_e,& N_e\rightarrow PN_e, \\
\phi_1' \rightarrow P^{-k}\phi_1', & \phi_2'\rightarrow P^{-k} \phi_2',\\
\phi_3' \rightarrow P^{-l}\phi_3', & \phi_4'\rightarrow P^{-l} \phi_4',\\
\phi_5' \rightarrow P^{-m}\phi_5', & \phi_6'\rightarrow P^{-m} \phi_6',\\
\phi_1\rightarrow P^k\phi_1, & \phi_2\rightarrow P^k\phi_2 ,\\
\phi_3\rightarrow P^l\phi_3, & \phi_4\rightarrow P^l\phi_4,\\
\phi_5\rightarrow P^m\phi_5, & \phi_6\rightarrow P^m\phi_6,\\
\phi_9\rightarrow P^{-1}\phi_9, & \phi_{10}\rightarrow P\phi_{10},
\end{matrix}
\right.
\end{eqnarray}
where $k$, $l$, and $m$ are some integers. For the symmetry ${\mathcal D}_6$
we additionally require that the Froggatt--Nielsen states with non-zero
hypercharge can only be multiplied by factors $P^n$, where $n$ is an
integer multiple of $k$, $l$, or $m$, and that the differences
$|k-l|$, $|k-m|$, and $|l-m|$ are
sufficiently large. The only fermion with non-vanishing hypercharge
that transforms differently is the right-handed electron $E_e$.
These symmetries restrict the allowed combinations of the 
scalar fields in the higher-dimensional lepton mass operators as well
as in the renormalizable terms of the multi-scalar potential. Thus,
possibly dangerous terms in the multi-scalar potential can be
forbidden, which could otherwise spoil the vacuum alignment mechanism.

At this stage, it is appropriate to point out some implications concerning
the nature of the discrete symmetries. It has been found
that relations between Yukawa couplings established by standard discrete
symmetries can only remain unbroken by quantum gravity corrections if the
discrete symmetries are gauged \cite{Krauss:1989zc}. These ``discrete
gauge symmetries'' appear in continuum theories when a gauge symmetry group
$G$ is broken to a discrete symmetry subgroup $H$. Since acceptable continuous
gauge theories have to be free from chiral anomalies, one obtains
discrete anomaly cancellation conditions, which strongly constrain the
massless fermion content of the theory \cite{Ibanez:1991hv}. At a more
fundamental level, this implies for our model that the permutation symmetries
must actually be gauged, since the relations, which are
based on them, will prove to be crucial for obtaining essentially
strict maximal atmospheric mixing. However, it is interesting to note
that in $M$-theory, one expects the discrete symmetries to be
always anomaly-free \cite{Witten:2001bf}.

\section{The multi-scalar potential}
\label{sec:scalar_pot}

\subsection{The two-Higgs doublet potential}

{}From the most general two-Higgs doublet potential of the fields
$H_1$ and $H_2$ 
(for an extensive review on electroweak Higgs potentials, see, \eg,
\Ref\cite{Sher:1989mj}) we conclude that, in presence of the
${\mathbb{Z}}_2$ symmetry, which distinguishes between $H_1$ and $H_2$ (see
\Sec\ref{sec:rep_con}), each of these fields appears only to the
second or fourth power in the multi-scalar potential. Hence, in any
renormalizable terms of the multi-scalar potential, which mix $H_1$ or
$H_2$ with the SM singlet scalar fields, the Higgs doublets are only allowed to
appear in terms of their absolute squares $|H_1|^2$ and $|H_2|^2$.
Next, since the Higgs doublets carry zero $Q_1,Q_2,$ and $Q_3$ charges
and are ${\mathcal{D}}_i$-singlets, where  $i=1,2,\ldots,6$, there exists a
range of parameters in the multi-scalar potential for which the standard
two-Higgs electroweak symmetry breaking is possible. Furthermore,
this implies that we can, without loss of generality, separate the SM
singlet scalar part from the Higgs-doublet part in the multi-scalar
potential by formally absorbing the absolute squares of the VEVs
$|\langle H_1\rangle|^2$ and $|\langle H_2\rangle|^2$ into the
coupling constants of the
mixed terms. Then, since the vacuum alignment mechanism of the SM singlet
fields is independent from the details of the Higgs doublet physics, we can in
what follows discard the effects of the Higgs doublets and fully
concentrate on the properties of the SM singlet scalar fields.

\subsection{Interactions of the fields $\phi_9$ and $\phi_{10}$}
\label{sec:phi9and10}

The symmetry
${\mathcal{D}}_6$ requires that the fields $\phi_9$ and $\phi_{10}$
enter the renormalizable interactions of the scalar fields only in
terms of the operators 
$\phi_9\phi_{10},\phi_9^\dagger\phi_{10}^\dagger,|\phi_9|^2$, and
$|\phi_{10}|^2$. The product
$\phi_9\phi_{10}$ has the U(1) charge structure $(-2,0,1)$, implying that the
only renormalizable interaction in the scalar potential, which involves this
product, is actually $\sim |\phi_9|^2|\phi_{10}|^2$. We can assume
that both of the fields $\phi_9$ and $\phi_{10}$ finally develop non-vanishing
VEVs with magnitudes $|\langle\phi_9\rangle|$ and
$|\langle\phi_{10}\rangle|$ that are not too large.
Since the fields $\phi_9$ and $\phi_{10}$ are singlets under
transformations of all the
permutation symmetries ${\mathcal{D}}_2,{\mathcal{D}}_3$, and
${\mathcal{D}}_4$, 
they will have no effect on the relative alignment of
the rest of the scalar fields, because they enter the corresponding
interactions 
only in terms of the absolute squares $|\langle\phi_9\rangle|^2$ and
$|\langle\phi_{10}\rangle|^2$. For this reason we will, without loss
of generality, discard the terms in the scalar potential
which involve the fields $\phi_9$ and $\phi_{10}$ in our considerations.

{}From the assignment of the U(1) charges $Q_1,Q_2$, and $Q_3$ and the
symmetry ${\mathcal{D}}_6$ (which does not permute any of the fields)
it follows that any renormalizable
term in the scalar potential which involves the SM singlet fields
$\phi_1,\phi_2,\phi_5,\phi_6,\phi_7,\phi_8$, or $\theta$ can
only be allowed if these fields appear in one of the following combinations:
\begin{equation}\label{eq:2fold}
\phi_{1,2}\phi_{1,2}^\dagger\:,\quad \phi_{5,6}\phi_{5,6}^\dagger\:,\quad
\phi_{5,6}\theta\:,\quad\phi_{7,8}\phi_{7,8}^\dagger\:,\quad|\theta|^2\:,
\end{equation}
where ``$\phi_{i,j}$'' ($i,j=1,2,\dots,8$) denotes some linear
combination of the 
fields $\phi_i$ and $\phi_j$. Among the products in
\eq~(\ref{eq:2fold}) only the 
product $\phi_{5,6}\theta$ transforms non-trivially under the symmetry
${\mathcal{D}}_1$. In addition, since the fields
$\phi_1',\phi_2',\ldots,\phi_6'$ are
${\mathcal{D}}_1$-singlets, we observe that the fields $\phi_3$ and
$\phi_4$, which carry a non-zero ${\mathcal{D}}_1$-charge, can only
appear either in the combination $\phi_{3,4}\phi_{3,4}^\dagger$ or in
the combination $\phi_{3,4}^\dagger\phi_{5,6}\theta$. However, the latter
combination is forbidden by $\mathcal{D}_6$-invariance. 

\subsection{Interactions of the field $\theta$}
\label{sec:theta}

As in \Sec\ref{sec:phi9and10}, we can assume that the field
$\theta$ finally develops a non-vanishing VEV with a magnitude
$|\langle \theta\rangle|$ that is not too large. Except for the
combination $\phi_{5,6}\theta$ in 
\eq~(\ref{eq:2fold}) all scalar interactions involve an equal
number (0, 1, or 2)
of the fields $\theta$ and its adjoint $\theta^\dagger$, which can then be
paired to the absolute square $|\theta|^2$. Since the interaction
$\phi_{5,6}\theta|\theta|^2$ is forbidden by the symmetries
${\mathcal{D}}_1$ and ${\mathcal{D}}_6$ and
$\theta$ is a total singlet under transformations of the discrete symmetries
${\mathcal{D}}_1,{\mathcal{D}}_2,\ldots,{\mathcal{D}}_6$, all
terms which involve the absolute square $|\theta|^2$ will have no influence on
the relative alignment of the rest of the scalar fields. For this
reason we can, without loss of generality, omit the terms in the
scalar potential which involve $|\theta|^2$ in our considerations.

\subsection{Interactions of the fields $\phi_7$ and $\phi_8$}
\label{sec:phi7and8}

{}From the U(1) charge assignment we conclude that only an even number of the
fields $\phi_7$ and $\phi_8$ (or their complex conjugates) can participate in
the scalar interactions. Let us now especially consider the operator
$\phi_7^\dagger\phi_8$ (or equivalently its complex
conjugate). Under application of each of the symmetries ${\mathcal{D}}_2$ and
${\mathcal{D}}_3$ the operator $\phi_7^\dagger\phi_8$ changes sign.
Therefore, the symmetry ${\mathcal{D}}_2$ requires this
operator to couple to one of the fields taken from the
set $\{\phi_1',\phi_2',\phi_1,\phi_2,\phi_7\}$. Simultaneously, the
symmetry ${\mathcal{D}}_3$ requires this operator to
couple only to fields taken from the extended set
$\{\phi_3',\phi_4',\phi_5',\phi_6',\phi_3,\phi_4,\phi_5,\phi_6,\phi_7\}$.
Both conditions can only be fulfilled if the operator
$\phi_7^\dagger\phi_8$ couples to the field
$\phi_7$ or its complex conjugate, but not to the operator products
$(\phi_7)^2,(\phi_7^\dagger)^2$, or $|\phi_7|^2$. Since the product
$\phi_7^\dagger\phi_8$ carries
the U(1) charges $(0,0,0)$, the operator $\phi_7^\dagger\phi_8$ can only couple
to some linear combination of the operators $\phi_7\phi_8^\dagger$ and
$\phi_8\phi_7^\dagger$.
As a consequence, if a general operator $\phi_{7,8}\phi_{7,8}^\dagger$
enters an interaction with scalars, which are different from the
fields $\phi_7$ and $\phi_8$, then this operator
$\phi_{7,8}\phi_{7,8}^\dagger$ is a linear combination of the absolute
squares $|\phi_7|^2$ and $|\phi_8|^2$.

Let us denote by $\phi_i$ and $\phi_j$ two scalar fields, which are different
from the
fields $\phi_7$ and $\phi_8$. Taking \eq~(\ref{eq:2fold}) and the operator
$\phi_{3,4}\phi_{3,4}^\dagger$ into account, the operator product
$\phi_i\phi_j$ must be of one of the following types:
\begin{equation}\label{eq:phiiphij}
 \phi_{1,2}\phi_{1,2}^\dagger\:,\quad\phi_{3,4}\phi_{3,4}^\dagger\:,\quad
 \phi_{5,6}\phi_{5,6}^\dagger\:,\quad\phi_{1,2}'{\phi'}^\dagger_{1,2}\:,\quad
 \phi_{3,4}'{\phi'}_{3,4}^\dagger\:,\quad\phi_{5,6}'{\phi'}_{5,6}^\dagger\:,
\end{equation}
where the last three combinations follow from the
${\mathcal{D}}_6$-invariance. Next, the symmetry
${\mathcal{D}}_5$ implies that $\phi_j=\phi_i^\dagger$ for the
combinations in \eq~(\ref{eq:phiiphij}), \ie, the product is the
absolute square $\phi_i\phi_j=|\phi_i|^2$. Using the result of the
previous paragraph, invariance under transformation of the symmetry
${\mathcal{D}}_4$ gives for the most general
interactions of the fields $\phi_7$ and $\phi_8$ with the other scalar fields
the terms
\begin{equation}
(|\phi_7|^2+|\phi_8|^2)\sum_{\varphi_i\neq \phi_7,\phi_8}c_i |\varphi_i|^2,
\end{equation}
where $\varphi_i$ can be any of the scalar fields,
which are not identical with the fields
$\phi_7$ or $\phi_8$ and $c_i$ are some real-valued coupling constants.
(Dimension-three terms of the types $|\phi_7|^2\varphi_i$ or
$|\phi_8|^2\varphi_i$, where $\varphi_i\neq\phi_7,\phi_8$, are forbidden by the
U(1) charge assignment and the symmetry ${\mathcal{D}}_6$,
which does not permute any fields.)
Taking everything into account,  the U(1) charge assignment and the
symmetry ${\mathcal{D}}_4$ restrict the most general terms in the
scalar potential, involving the fields $\phi_7$ and $\phi_8$, to be
\begin{eqnarray}\label{eq:V78}
V_{7,8} &=& \mu^2 (|\phi_7|^2+|\phi_8|^2) + \kappa (|\phi_7|^2+|\phi_8|^2)^2
+ (|\phi_7|^2+|\phi_8|^2) \sum_{\varphi_i\neq \phi_7,\phi_8}c_i
|\varphi_i|^2 \nonumber\\
&+& a \left[(\phi_7^\dagger\phi_8)^2 + (\phi_8^\dagger\phi_7)^2\right]
     + b|\phi_7|^2|\phi_8|^2,
\end{eqnarray}
where $\mu^2$, $\kappa$, $a$, and $b$ are real-valued constants.
Parameterizing the VEVs of $\phi_7$ and $\phi_8$ as
\begin{equation}
\left( \begin{array}{c} \langle \phi_7 \rangle \\ \langle \phi_8 \rangle
  \end{array} \right) = v \left( \begin{array}{c} {\rm e}^{{\rm i}
    \beta_1} \cos \alpha \\ {\rm e}^{{\rm i} \beta_2} \sin \alpha
    \end{array} \right),
\end{equation}
where $v$ is some real-valued number, $\alpha$ is the angle which
rotates $\langle \phi_7 \rangle$ and $\langle \phi_8 \rangle$, and
$\beta_1$ and $\beta_2$ 
denote the phases of the VEVs, we observe in \eq~(\ref{eq:V78}) that
the first three terms exhibit an accidental ${\rm U(1)}_\alpha$
symmetry. This symmetry is broken by the last two terms, which are
therefore responsible for the alignment of the VEVs. Note that the
scalar potential is symmetric under the exchange
$\phi_7\leftrightarrow\phi_8$ and that the
term with coefficient $a$ is $\sim \Re \left[ (\phi_7^\dagger
\phi_8)^2 \right]$. In \eq~(\ref{eq:V78}), we will choose
$\kappa > 0$ and assume the rest of the coupling constants to be
negative. Then, the lowest energy state is characterized by $\alpha =
\tfrac{\pi}{4}$ and $\beta_1 - \beta_2 \in \{0,\pi\}$ or equivalently
\begin{equation}\label{eq:VEVsV78}
\langle\phi_7\rangle=\pm\langle\phi_8\rangle,
\end{equation}
\ie, the VEVs are degenerate up to a sign. When considering the Yukawa
interactions of the neutrinos, it will turn out that the degeneracy of the
VEVs is responsible for a nearly maximal atmospheric mixing angle. 
Thus, we can from now on restrict our discussion of the
scalar potential to the fields $\phi_1,\phi_2,\ldots,\phi_6$, and
$\phi_1',\phi_2',\ldots ,\phi_6'$. This discussion will follow in the three
coming subsections.

\subsection{The potential of the fields $\phi_1,\phi_2,\ldots ,\phi_6$}

In all two-fold and four-fold products involving only the fields $\phi_i$
($i=1,2,\ldots ,6$), the discrete symmetry $\mathcal{D}_5$ requires
the number of these fields, which are denoted by even
(or odd) indices, to be even. In the scalar potential, linear and tri-linear
terms of the fields $\phi_1,\phi_2,\ldots,\phi_6$ are forbidden by the symmetry
${\mathcal{D}}_6$. Taking the combinations in \eq~(\ref{eq:2fold})
and $\phi_{3,4}\phi_{3,4}^\dagger$ into account, the allowed two-fold products
of the fields $\phi_1,\phi_2,\ldots,\phi_6$ can only be of the type
$|\phi_i|^2$, \ie, they must be absolute squares of the fields.
Similarly, we obtain that all four-fold products of the fields
$\phi_1,\phi_2,\ldots,\phi_6$ must be of the types
\begin{eqnarray}\label{eq:4fold}
 &&(\phi_1\phi_2^\dagger)^2\:,\quad
    \phi_1\phi_2^\dagger\phi_3\phi_4^\dagger\:,\quad
	\phi_1\phi_2^\dagger\phi_4\phi_3^\dagger\:,\quad
	\phi_1\phi_2^\dagger\phi_5\phi_6^\dagger\:,\quad
    \phi_1\phi_2^\dagger\phi_6\phi_5^\dagger\:,\nonumber\\
 &&(\phi_3\phi_4^\dagger)^2\:,\quad
    \phi_3\phi_4^\dagger\phi_5\phi_6^\dagger\:,\quad
	\phi_3\phi_4^\dagger\phi_6\phi_5^\dagger\:,\quad
   (\phi_5\phi_6^\dagger)^2\:,\quad
    |\phi_i|^2|\phi_j|^2\:,\quad
	|\phi_i|^4\:,
\end{eqnarray}
and their complex conjugates, where $i,j=1,2,\ldots,6$.
A general four-fold product of the types in \eq~(\ref{eq:4fold}) can
be written as
\begin{eqnarray}\label{eq:general4fold}
  &  &(a\phi_i^\dagger\phi_j+b\phi_j^\dagger\phi_i)\phi_k^\dagger\phi_l+
 (c\phi_i^\dagger\phi_j+d\phi_j^\dagger\phi_i)\phi_l^\dagger\phi_k+{\rm
 h.c.}\nonumber\\
 &=& \left[(a+d^\ast)\phi_i^\dagger\phi_j+(b+c^\ast)\phi_j^\dagger\phi_i\right]
 \phi_k^\dagger\phi_l
 +\left[(c+b^\ast)\phi_i^\dagger\phi_j+(d+a^\ast)\phi_j^\dagger\phi_i\right]
 \phi_l^\dagger\phi_k,\nonumber\\
\end{eqnarray}
where $a,b,c$, and $d$ are complex-valued constants.
Assume that $i\neq j$. Then, from \eq~(\ref{eq:4fold}) we observe that
$k\neq l$ and we can, without loss of generality, assume that the
index-pairs $(i,j)$ and $(k,l)$, respectively,
combine the fields which are interchanged by the discrete symmetry
${\mathcal{D}}_2$ or ${\mathcal{D}}_3$. (If $i=j$, then it follows
from \eq~(\ref{eq:4fold}) that $k=l$, which will be discussed below.)
Let, in addition, $\{ i,j\}\neq \{k,l\}$. Then, application of the symmetries
${\mathcal{D}}_2$ and ${\mathcal{D}}_3$ yields $a+d^\ast=d+a^\ast$ and
$b+c^\ast=c+b^\ast$.
 We can therefore rename the constants as
$a+d^\ast\rightarrow a$ and $b+c^\ast\rightarrow b$, where now $a$ and $b$ are
real constants, and write the term in
\eq~(\ref{eq:general4fold}) as
\begin{eqnarray}\label{eq:realandimaginary}
 &&(a\phi_i^\dagger\phi_j+b\phi_j^\dagger\phi_i)\phi_k^\dagger\phi_l+
 (b\phi_i^\dagger\phi_j+a\phi_j^\dagger\phi_i)\phi_l^\dagger\phi_k\nonumber\\
&=&\frac{a+b}{2}(\phi_i^\dagger\phi_j+\phi_j^\dagger\phi_i)
(\phi_k^\dagger\phi_l+\phi_l^\dagger\phi_k)
+
\frac{a-b}{2}(\phi_i^\dagger\phi_j-\phi_j^\dagger\phi_i)
(\phi_k^\dagger\phi_l-\phi_l^\dagger\phi_k)\nonumber\\
&=&\frac{a+b}{2}\:\Re(\phi_i^\dagger\phi_j)\:\Re(\phi_k^\dagger\phi_l)
 -\frac{a-b}{2}\:\Im(\phi_i^\dagger\phi_j)\:\Im(\phi_k^\dagger\phi_l).
\end{eqnarray}
Since the fields $\phi_3,\phi_4,\phi_5$, and $\phi_6$ are singlets
under transformation of the discrete symmetry ${\mathcal{D}}_2$, we
can have $a\neq b$ in the case that $(\phi_i,\phi_j)=(\phi_3,\phi_4)$ and
$(\phi_k,\phi_l)=(\phi_5,\phi_6)$. However, if
$(i,j)=(1,2)$, then application of the discrete symmetry ${\mathcal{D}}_2$
further constrains the constants in the above general form to fulfill
$a=b$, and therefore, the last term in \eq~(\ref{eq:realandimaginary})
vanishes.

As a cause of the symmetries ${\mathcal{D}}_2$ and ${\mathcal{D}}_3$,
the products $(\phi_i\phi_j^\dagger)^2$ in \eq~(\ref{eq:4fold}),
where $(i,j)=(1,2),(3,4),(5,6)$, appear in the potential always as
\begin{equation}
a\left[(\phi_i\phi_j^\dagger)^2+(\phi_j^\dagger\phi_i)^2\right] =
2 a\Re \left[ (\phi_i\phi_j^\dagger)^2 \right],
\end{equation}
where $a$ is some real-valued constant.

Let us now turn the discussion to the terms
$|\phi_i|^2|\phi_k|^2$ in \eq~(\ref{eq:4fold}), where $i\neq
k$. Assume that the fields $\phi_i$ and $\phi_k$ cannot be combined
into one of the pairs $(\phi_1,\phi_2),(\phi_3,\phi_4)$, or
$(\phi_5,\phi_6)$. Then, a general term of this type is on the form
\begin{equation}
 (a|\phi_i|^2+b|\phi_j|^2)|\phi_k|^2+(c|\phi_i|^2+d|\phi_j|^2)|\phi_l|^2,
\end{equation}
where $a,b,c$, and $d$ are real-valued constants and
\newline
$(\phi_i,\phi_j),(\phi_k,\phi_l)\in\{(1,2),(3,4),(5,6)\}$. Application of the
symmetries ${\mathcal{D}}_2$ and ${\mathcal{D}}_3$ yields the conditions $a=d$
and $b=c$, and thus, we can rewrite the above part of the potential as
\begin{equation}\label{eq:absolutsquares}
\frac{a+b}{2}(|\phi_i|^2+|\phi_j|^2)(|\phi_k|^2+|\phi_l|^2)
+ \frac{a-b}{2}(|\phi_i|^2-|\phi_j|^2)(|\phi_k|^2-|\phi_l|^2).
\end{equation}
If $(i,j)=(3,4)$ and $(k,l)=(5,6)$, then in general $a\neq b$, since the fields
$\phi_3,\phi_4,\phi_5$, and $\phi_6$ are
${\mathcal{D}}_2$-singlets. However, if $(i,j)=(1,2)$, then $a=b$ and
the part in \eq~(\ref{eq:absolutsquares}) which is proportional to
$(a-b)/2$ vanishes.

If in the combination $|\phi_i|^2|\phi_j|^2$ in \eq~(\ref{eq:4fold}) the fields
$(\phi_i,\phi_j)$ form one of the pairs $(\phi_1,\phi_2),(\phi_3,\phi_4)$,
or $(\phi_5,\phi_6)$, then the combination $|\phi_i|^2|\phi_j|^2$ is a total
singlet (on its own) and it can be written directly into the scalar
potential as $a|\phi_i|^2|\phi_j|^2$, where $a$ is some real-valued constant.

Moreover, the symmetries ${\mathcal{D}}_2$ and ${\mathcal{D}}_3$
enforce the products $|\phi_i|^2$ and $|\phi_i|^4$ [in 
\eq~(\ref{eq:4fold})] to appear in the scalar potential only as
\begin{eqnarray}
&&\mu_1^2 \left( |\phi_1|^2 + |\phi_2|^2 \right)
+ \mu_2^2 \left( |\phi_3|^2 + |\phi_4|^2 \right)
+ \mu_3^2 \left( |\phi_5|^2 + |\phi_6|^2 \right)\nonumber\\
&+& \kappa_1 \left( |\phi_1|^4 + |\phi_2|^4\right)
  + \kappa_2 \left( |\phi_3|^4 + |\phi_4|^4 \right)
  + \kappa_3 \left( |\phi_5|^4 + |\phi_6|^4 \right),
\end{eqnarray}
where $\mu_1^2$, $\mu_2^2$, $\mu_3^2$, $\kappa_1$, $\kappa_2$, and
$\kappa_3$ are real-valued constants. In total, the most general scalar
potential involving only the unprimed fields, but neither the fields
$\phi_7$, $\phi_8$, $\phi_9$, and $\phi_{10}$ nor the product
$|\theta|^2$, reads
\begin{eqnarray}
V_1 &=& \mu_1^2 \left( |\phi_1|^2 + |\phi_2|^2 \right)
+ \mu_2^2 \left( |\phi_3|^2 + |\phi_4|^2 \right)
+ \mu_3^2 \left( |\phi_5|^2 + |\phi_6|^2 \right)\nonumber\\
&+& \kappa_1 \left( |\phi_1|^2 + |\phi_2|^2 \right)^2
+ \kappa_2 \left( |\phi_3|^2 + |\phi_4|^2 \right)^2
+ \kappa_3 \left( |\phi_5|^2 + |\phi_6|^2 \right)^2
 \nonumber\\
&+& d_1 \left( |\phi_1|^2 + |\phi_2|^2 \right) \left( |\phi_3|^2 +
 |\phi_4|^2 \right) + d_2 \left( |\phi_1|^2 + |\phi_2|^2 \right)
 \left( |\phi_5|^2 + |\phi_6|^2 \right) \nonumber\\ 
&+& d_3 \left( |\phi_3|^2 + |\phi_4|^2 \right) \left( |\phi_5|^2 +
 |\phi_6|^2 \right) + d_4 \left( |\phi_3|^2 - |\phi_4|^2 \right)
 \left( |\phi_5|^2 - |\phi_6|^2 \right) \nonumber\\
&+& d_5 |\phi_1^\dagger \phi_2|^2 + d_6 |\phi_3^\dagger \phi_4|^2
+ d_7|\phi_5^\dagger \phi_6|^2 + d_8 \left[ (\phi_1^\dagger \phi_2)^2
+ (\phi_2^\dagger \phi_1)^2 \right] \nonumber\\
&+& d_9 \left[(\phi_3^\dagger \phi_4)^2+(\phi_4^\dagger \phi_3)^2 \right]
 + d_{10} \left[(\phi_5^\dagger \phi_6)^2 +
(\phi_6^\dagger \phi_5)^2 \right]\nonumber\\
&+& d_{11} \left(\phi_1^\dagger \phi_2 + \phi_2^\dagger \phi_1\right)
\left(\phi_3^\dagger \phi_4 + \phi_4^\dagger \phi_3\right) \nonumber\\
&+& d_{12} \left(\phi_1^\dagger \phi_2 + \phi_2^\dagger \phi_1\right)
\left(\phi_5^\dagger \phi_6 + \phi_6^\dagger \phi_5\right) \nonumber\\
&+& d_{13} \left(\phi_3^\dagger \phi_4 + \phi_4^\dagger \phi_3\right)
\left(\phi_5^\dagger \phi_6 + \phi_6^\dagger \phi_5\right) \nonumber\\
&+& d_{14} \left(\phi_3^\dagger \phi_4 - \phi_4^\dagger \phi_3\right)
\left(\phi_5^\dagger \phi_6 - \phi_6^\dagger \phi_5\right), \nonumber\\
\label{eq:V1}
\end{eqnarray}
where $d_1,d_2,\ldots,d_{14}$ are real-valued constants. We will assume that
$\kappa_1,\kappa_2,\kappa_3 > 0$ and $d_{11},d_{13} > 0$ and we will choose all
other coupling constants to be negative. Again, we observe that the
first nine terms in \eq~(\ref{eq:V1}) exhibit three accidental U(1)
symmetries, which act on the pairs of VEVs $(\langle \phi_1
\rangle,\langle \phi_2 \rangle)$, $(\langle \phi_3 \rangle,\langle
\phi_4 \rangle)$, and $(\langle \phi_5 \rangle,\langle \phi_6
\rangle)$, respectively. The rest of the terms in the potential $V_1$
break these symmetries and will therefore determine the vacuum
alignment mechanism of the fields. First, we note that the term with the
coefficient $d_4$ tends (for large values of $|d_4|$) to induce a
splitting between $|\langle \phi_3 \rangle|$ and $|\langle \phi_4
\rangle|$ as well as between $|\langle \phi_5 \rangle|$ and $|\langle
\phi_6 \rangle|$. Second, we observe that the term with the
coefficient $d_{14}$ has (for large values of $|d_{14}|$) the tendency
to trigger relative phases (different from $0$ and $\pi$) between
$\langle \phi_3 \rangle$ and $\langle \phi_4 \rangle$ as well as
between $\langle \phi_5 \rangle$ and $\langle \phi_6
\rangle$. However, if we require that
\begin{equation}
d_6 d_7 > 4 d_{4}^2 \quad \mbox{and} \quad d_9
d_{10} > 4 d_{14}^2,
\end{equation}
then the potential $V_1$ is minimized by the VEVs of the unprimed
fields, which are pairwise degenerate in their magnitudes, \ie, they satisfy
\begin{subequations}\label{eq:VEVsV3}
\begin{equation}
 |\langle\phi_1\rangle|=|\langle\phi_2\rangle|,\quad
 |\langle\phi_3\rangle|=|\langle\phi_4\rangle|,\quad
 |\langle\phi_5\rangle|=|\langle\phi_6\rangle|
\end{equation}
and are also pairwise relatively real, \ie,
\begin{equation}
 \frac{\langle\phi_1\rangle}{\langle\phi_2\rangle},
 \frac{\langle\phi_3\rangle}{\langle\phi_4\rangle},
 \frac{\langle\phi_5\rangle}{\langle\phi_6\rangle}
 \in\{-1,1\},
\end{equation}
where, in addition, the choice $d_{12} < 0$ and $d_{11},d_{13} > 0$
implies a correlation between the different pairs of VEVs in terms of
\begin{equation}
 \frac{\langle\phi_1\rangle}{\langle\phi_2\rangle}
 \frac{\langle\phi_5\rangle}{\langle\phi_6\rangle}=1\quad \mbox{and} \quad
 \frac{\langle\phi_1\rangle}{\langle\phi_2\rangle}
 \frac{\langle\phi_3\rangle}{\langle\phi_4\rangle}=-1,
\end{equation}
\end{subequations}
\ie, the relative sign between $\langle\phi_1\rangle$ and
$\langle\phi_2\rangle$ is equal to the relative sign between
$\langle\phi_5\rangle$ and $\langle\phi_6\rangle$ and opposite to the relative
sign between $\langle\phi_3\rangle$ and $\langle\phi_4\rangle$.

\subsection{The potential of the fields $\phi_1',\phi_2',\ldots,\phi_6'$}
\label{sec:V1}

The only two-fold products of the primed fields $\phi_i'$ ($i=1,2,\ldots,6$)
which are allowed by the discrete symmetries
${\mathcal{D}}_1,{\mathcal{D}}_5$, and ${\mathcal{D}}_6$ are the absolute
squares $|\phi_i'|^2$. Furthermore, the permutation symmetries
${\mathcal{D}}_2$ and ${\mathcal{D}}_3$ yield for the most general
dimension-two terms of the primed fields the expression
\begin{equation}
\mu_4^2 (|\phi_1'|^2+|\phi_2'|^2) + 
\mu_5^2 (|\phi_3'|^2+|\phi_4'|^2) +
\mu_6^2 (|\phi_5'|^2+|\phi_6'|^2),
\end{equation}
where $\mu_4^2$, $\mu_5^2$, and $\mu_6^2$ are real-valued constants. The
permutation symmetries ${\mathcal{D}}_2$ and ${\mathcal{D}}_3$
additionally require the most general products of the absolute squares
$|\phi'_i|^2$ to be
\begin{eqnarray}
 &&\kappa_4 \left(|\phi_1'|^2+|\phi_2'|^2\right)^2 + \kappa_5
 \left(|\phi_3'|^2+|\phi_4'|^2\right)^2 + \kappa_6
 \left(|\phi_5'|^2+|\phi_6'|^2\right)^2 \nonumber\\
 &+& a_1
 \left(|\phi_1'|^2+|\phi_2'|^2\right)\left(|\phi_3'|^2+|\phi_4'|^2\right)
 + a_2 \left( |\phi_1'|^2 + |\phi_2'|^2 \right)
 \left( |\phi_5'|^2 + |\phi_6'|^2 \right) \nonumber\\ 
&+& a_3 \left( |\phi_3'|^2 + |\phi_4'|^2 \right) \left( |\phi_5'|^2 +
 |\phi_6'|^2 \right) + a_4 \left( |\phi_3'|^2 - |\phi_4'|^2 \right)
 \left( |\phi_5'|^2 - |\phi_6'|^2 \right) \nonumber\\
&+& a_5 |{\phi_1'}^\dagger \phi_2'|^2 + a_6 |{\phi_3'}^\dagger \phi_4'|^2
+ a_7|{\phi_5'}^\dagger \phi_6'|^2,
\end{eqnarray}
where $\kappa_4$, $\kappa_5$, $\kappa_6$, and $a_1,a_2,\ldots,a_7$ are
real-valued constants. Note that three-fold products of the primed fields are
forbidden by the ${\mathcal{D}}_6$-charge assignment. The discrete
symmetry ${\mathcal{D}}_6$ requires
that the remaining interactions of the primed fields can be written as products
of the operators ${\phi'}_1^\dagger\phi'_2$,
${\phi'}_3^\dagger\phi'_4$, and ${\phi'}_5^\dagger\phi'_6$
(and their complex conjugates). Hence, the operators which involve the
fields of only one of the pairs $(\phi_1',\phi_2')$,
$(\phi_3',\phi_4')$, or
$(\phi_5',\phi_6')$ are restricted by the symmetries ${\mathcal{D}}_3$
and ${\mathcal{D}}_4$ to be on the form
\begin{eqnarray}
&& a_8 \left[({\phi_1'}^\dagger \phi_2')^2+({\phi_2'}^\dagger
{\phi_1'})^2\right] + a_9 \left[({\phi_3'}^\dagger
\phi_4')^2+({\phi_4'}^\dagger \phi_3')^2\right] \nonumber\\
&+& a_{10}
  \left[({\phi_5'}^\dagger \phi_6')^2+({\phi_6'}^\dagger
    \phi_5')^2\right] \nonumber\\
&=& 2 a_8 \Re \left[ ({\phi_1'}^\dagger {\phi_2'})^2
\right] + 2 a_9 \Re \left[ ({\phi_3'}^\dagger \phi_4')^2\right] + 2
  a_{10} \Re \left[ ({\phi_5'}^\dagger \phi_6')^2\right],
\end{eqnarray}
where $a_8$, $a_9$, and $a_{10}$ are real-valued constants. Furthermore, the
symmetries ${\mathcal{D}}_2,{\mathcal{D}}_3,{\mathcal{D}}_5$, and 
${\mathcal{D}}_6$ restrict the only dimension-four terms involving at
least three different primed fields (or their complex conjugates) to be
\begin{eqnarray}
&& a_{11} \left({\phi_1'}^\dagger \phi'_2 + {\phi_2'}^\dagger \phi_1'\right)
\left({\phi_3'}^\dagger \phi_4' + {\phi_4'}^\dagger \phi_3'\right)
\nonumber\\
&+& a_{12} \left({\phi_1'}^\dagger \phi_2' + {\phi_2'}^\dagger {\phi_1'}\right)
\left({\phi_5'}^\dagger {\phi_6'} + {\phi_6'}^\dagger {\phi_5'}\right)
\nonumber\\
&+& a_{13} \left({\phi_3'}^\dagger {\phi_4'} + {\phi_4'}^\dagger
    {\phi_3'}\right)  
\left({\phi_5'}^\dagger {\phi_6'} + {\phi_6'}^\dagger {\phi_5'}\right)
\nonumber\\
&+& a_{14} \left({\phi_3'}^\dagger {\phi_4'} - {\phi_4'}^\dagger
    {\phi_3'}\right) 
\left({\phi_5'}^\dagger {\phi_6'} - {\phi_6'}^\dagger {\phi_5'}\right)
\nonumber\\ &=& 4 a_{11} \Re ({\phi_1'}^\dagger \phi_2') \Re
({\phi_3'}^\dagger \phi_4') 
+ 4 a_{12} \Re ({\phi_1'}^\dagger \phi_2') \Re ({\phi_5'}^\dagger
\phi_6') \nonumber\\
&+& 4 a_{13} \Re ({\phi_3'}^\dagger \phi_4') \Re ({\phi_5'}^\dagger
\phi_6')
- 4 a_{14} \Im ({\phi_3'}^\dagger \phi_4') \Im ({\phi_5'}^\dagger \phi_6'),
\end{eqnarray}
where $a_{11}$, $a_{12}$, $a_{13}$, and $a_{14}$ are real-valued
constants. Taking everything
into account, the most general scalar potential involving only the
primed fields $\phi_i'$ reads
\begin{eqnarray}\label{eq:V1'}
V_1' &=&
\mu_4^2 (|\phi_1'|^2+|\phi_2'|^2) + 
\mu_5^2 (|\phi_3'|^2+|\phi_4'|^2) +
\mu_6^2 (|\phi_5'|^2+|\phi_6'|^2) \nonumber\\
 &+&\kappa_4 (|\phi_1'|^2+|\phi_2'|^2)^2 + \kappa_5
 (|\phi_3'|^2+|\phi_4'|^2)^2 + \kappa_6 (|\phi_5'|^2+|\phi_6'|^2)^2
 \nonumber\\
 &+& a_1 (|\phi_1'|^2+|\phi_2'|^2)(|\phi_3'|^2+|\phi_4'|^2) + a_2 \left( |\phi_1'|^2 + |\phi_2'|^2 \right)
 \left( |\phi_5'|^2 + |\phi_6'|^2 \right) \nonumber\\ 
&+& a_3 \left( |\phi_3'|^2 + |\phi_4'|^2 \right) \left( |\phi_5'|^2 +
 |\phi_6'|^2 \right) + a_4 \left( |\phi_3'|^2 - |\phi_4'|^2 \right)
 \left( |\phi_5'|^2 - |\phi_6'|^2 \right) \nonumber\\
&+& a_5 |{\phi_1'}^\dagger \phi_2'|^2 + a_6 |{\phi_3'}^\dagger \phi_4'|^2
+ a_7|{\phi_5'}^\dagger \phi_6'|^2
+ a_8 \left[({\phi_1'}^\dagger \phi_2')^2+({\phi_2'}^\dagger
{\phi_1'})^2\right] \nonumber\\
&+& a_9 \left[({\phi_3'}^\dagger
\phi_4')^2+({\phi_4'}^\dagger \phi_3')^2\right] + a_{10}
  \left[({\phi_5'}^\dagger \phi_6')^2+({\phi_6'}^\dagger
    \phi_5')^2\right] \nonumber\\
&+& a_{11} \left({\phi_1'}^\dagger \phi'_2 + {\phi_2'}^\dagger \phi_1'\right)
\left({\phi_3'}^\dagger \phi_4' + {\phi_4'}^\dagger \phi_3'\right)
\nonumber\\
&+& a_{12} \left({\phi_1'}^\dagger {\phi_2'} + {\phi_2'}^\dagger
    {\phi_1'}\right) 
\left({\phi_5'}^\dagger {\phi_6'} + {\phi_6'}^\dagger {\phi_5'}\right)
\nonumber\\
&+& a_{13} \left({\phi_3'}^\dagger {\phi_4'} + {\phi_4'}^\dagger
    {\phi_3'}\right)
\left({\phi_5'}^\dagger {\phi_6'} + {\phi_6'}^\dagger {\phi_5'}\right)
\nonumber\\
&+& a_{14} \left({\phi_3'}^\dagger {\phi_4'} - {\phi_4'}^\dagger
    {\phi_3'}\right) \left({\phi_5'}^\dagger {\phi_6'} - {\phi_6'}^\dagger
     {\phi_5'}\right).
\end{eqnarray}
In \eq~(\ref{eq:V1'}), we will assume that
$\kappa_4,\kappa_5,\kappa_6 > 0$ and we will
choose all other coupling constants to be negative. As in the
discussion of the potential $V_1$, we observe that the first nine
terms in \eq~(\ref{eq:V1'}) exhibit three accidental U(1) symmetries,
which act on the pairs of VEVs $(\langle \phi_1' \rangle, \langle
\phi_2' \rangle)$, $(\langle \phi_3' \rangle, \langle
\phi_4' \rangle)$, and $(\langle \phi_5' \rangle, \langle
\phi_6' \rangle)$, respectively. The rest of the terms in the
potential $V_1'$ break these symmetries and will therefore determine the vacuum
alignment mechanism of the fields. If we, in analogy to the potential
$V_1$, require that
\begin{equation}
a_6 a_7 > 4 a_4^2 \quad \mbox{and} \quad a_9 a_{10} > 4 a_{14}^2,
\end{equation} 
then the potential $V_1'$ is minimized by the VEVs of the primed
fields, which are pairwise degenerate in their magnitudes, \ie, they satisfy
\begin{subequations}\label{eq:VEVsV1}
\begin{equation}
 |\langle\phi'_1\rangle|=|\langle\phi'_2\rangle|,\quad
 |\langle\phi'_3\rangle|=|\langle\phi'_4\rangle|,\quad
 |\langle\phi'_5\rangle|=|\langle\phi'_6\rangle|
\end{equation}
and are also pairwise relatively real obeying
\begin{equation}\label{eq:VEVsV1b}
 \frac{\langle\phi'_1\rangle}{\langle\phi'_2\rangle}=
 \frac{\langle\phi'_3\rangle}{\langle\phi'_4\rangle}=
 \frac{\langle\phi'_5\rangle}{\langle\phi'_6\rangle}
 \in\{-1,1\}.
\end{equation}
\end{subequations}
Note that in \eq~(\ref{eq:VEVsV1b}) the pairs of the VEVs
$(\langle\phi_1'\rangle,\langle\phi_2'\rangle)$,
$(\langle\phi_3'\rangle,\langle\phi_4'\rangle)$, and
$(\langle\phi_5'\rangle,\langle\phi_6'\rangle)$ have the same relative
phase, \ie, the VEVs in the
pairs are either all oriented parallel or all oriented antiparallel.

\subsection{Mixing among the fields $\phi_1',\phi_2',\ldots,\phi_6'$ and
$\phi_1,\phi_2,\ldots,\phi_6$}
\label{sec:phi'phi}

The discrete symmetry ${\mathcal{D}}_6$ requires all renormalizable terms
mixing the primed fields
$\phi_i'$ or ${\phi_i'}^\dagger$ ($i=1,2,\ldots,6$) with the unprimed fields
$\phi_j$ or $\phi_j^\dagger$ ($j=1,2,\ldots,6$) to have an even mass dimension.
Taking the combinations in \eq~(\ref{eq:2fold}) and the product
$\phi_{3,4}^\dagger\phi_{3,4}$ into account (which are all
${\mathcal{D}}_6$-singlets), we obtain that the
${\mathcal{D}}_1$-invariant operator products, which
mix the fields $\phi_i'$ and $\phi_j$, must be of the types
\begin{equation}\label{eq:mixed}
 {\phi'}^\dagger_{i_1}\phi'_{i_2}\phi^\dagger_{1,2}\phi_{1,2}\:,\quad
 {\phi'}^\dagger_{i_1}\phi'_{i_2}\phi^\dagger_{3,4}\phi_{3,4}\:,\quad
 {\phi'}^\dagger_{i_1}\phi'_{i_2}\phi^\dagger_{5,6}\phi_{5,6}\:,
\end{equation}
where $i_1,i_2=1,2,\ldots,6$. The symmetry ${\mathcal{D}}_4$, which
acts only on the 
unprimed fields, requires the combinations in \eq~(\ref{eq:mixed}) to
be on the form ${\phi'}^\dagger_{i_1}\phi'_{i_2}|\phi^\dagger_j|^2$, where
$j=1,2,\ldots,6$. Next, the symmetries ${\mathcal{D}}_5$ and ${\mathcal{D}}_6$
imply that the operators in \eq~(\ref{eq:mixed}) are all in fact
$\sim|\phi_i'|^2|\phi_j|^2$. As a result, the most general renormalizable 
interactions of the fields $\phi_1',\phi_2',\ldots,\phi_6'$ with the fields
$\phi_1,\phi_2,\ldots,\phi_6$ are
\begin{eqnarray}\label{eq:V2}
V_2 &=& \left(|\phi_1'|^2+|\phi_2'|^2\right) \nonumber\\
&\times& \left[ b_1\left(|\phi_1|^2+|\phi_2|^2\right)
+ b_2\left(|\phi_3|^2+|\phi_4|^2\right)
+ b_3\left(|\phi_5|^2+|\phi_6|^2\right)\right] \nonumber\\
&+& b_4\left(|\phi_1'|^2-|\phi_2'|^2\right)\left(|\phi_1|^2-|\phi_2|^2\right)
\nonumber\\
&+& \left(|\phi_3'|^2+|\phi_4'|^2\right) \nonumber\\
&\times& \left[b_5\left(|\phi_1|^2+|\phi_2|^2\right)
+ b_6\left(|\phi_3|^2+|\phi_4|^2\right)
+ b_7\left(|\phi_5|^2+|\phi_6|^2\right)\right] \nonumber\\
&+& \left(|\phi_3'|^2-|\phi_4'|^2\right)
\left[b_8\left(|\phi_3|^2-|\phi_4|^2\right)
+ b_9\left(|\phi_5|^2-|\phi_6|^2\right)\right] \nonumber\\
&+& \left(|\phi_5'|^2+|\phi_6'|^2\right) \nonumber\\
&\times& \left[b_{10}\left(|\phi_1|^2+|\phi_2|^2\right)
+ b_{11}\left(|\phi_3|^2+|\phi_4|^2\right)
+ b_{12}\left(|\phi_5|^2+|\phi_6|^2\right)\right]\nonumber\\
&+&\left(|\phi_5'|^2-|\phi_6'|^2\right)
\left[b_{13}\left(|\phi_3|^2-|\phi_4|^2\right)
+b_{14}\left(|\phi_5|^2-|\phi_6|^2\right)\right],
\end{eqnarray}
where $b_1,b_2,\ldots,b_{14}$ are real-valued constants. In
\eq~(\ref{eq:V2}), we will assume all coupling constants to be negative.

In order to recover the (same) vacuum alignment mechanism that is operative for
the potentials $V_1$, $V_1'$, and $V_{7,8}$ also for the full SM
singlet scalar potential $V \equiv V_1 + V_1' + V_2 + V_{7,8}$, we will
have to ensure that the mixed terms in the potential $V_2$ do not
induce a splitting between the pairwise degenerate magnitudes of the VEVs.
If we require the coupling constants in the potentials $V_1$, $V_1'$, and
$V_2$ to fulfill
\begin{equation}
a_5 d_5 > 4 b_4^2, \quad a_6 d_6 > 4 b_8^2, \quad a_6 d_7 > 4
b_9^2, \quad a_7 d_6 > 4 b_{13}^2, \quad a_7 d_7 > 4 b_{14}^2,
\end{equation}
then the total multi-scalar potential $V$ is indeed minimized by the
VEVs of \eqs~(\ref{eq:VEVsV78}), (\ref{eq:VEVsV3}), and (\ref{eq:VEVsV1}).

We will suppose that all of the SM singlet scalar fields break the
flavor symmetries by acquiring their VEVs at a high mass scale
(somewhat below the fundamental mass scale $M_1$), and thereby, giving
rise to a small expansion parameter
\begin{equation}
\epsilon \simeq \frac{\langle \phi_i \rangle}{M_1} \simeq
\frac{\langle \phi_j' \rangle}{M_1} \simeq
\frac{\langle\theta\rangle}{M_1} \simeq 10^{-1},
\end{equation}
where $i = 1,2,\ldots,10$ and $j = 1,2,\ldots,6$. Such small hierarchies
can arise from large hierarchies in supersymmetric theories when the
scalar fields acquire their VEVs along a ``D-flat'' direction
\cite{Witten:1981kv,Leurer:1994gy}.

In the next two sections, \Secs\ref{sec:Yukawa_cl} and
\ref{sec:Yukawa_n}, we will discuss the (effective) Yukawa
interactions of the leptons that will lead to the mass matrix textures
of the leptons, and eventually, to the masses and mixings of them,
which will be discussed in \Secs\ref{sec:massmix} and \ref{sec:leptonic}.

\section{Yukawa interactions of the charged leptons}
\label{sec:Yukawa_cl}

\begin{figure}[ht!]
\begin{center}
\includegraphics*[bb = 190 520 430 670]{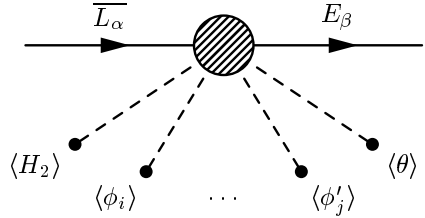}
\end{center}
\caption{Non-renormalizable terms generating the effective
Yukawa couplings in the matrix ${(\mathcal{O}}^\ell_{\alpha\beta})$.}
\label{fig:effective}
\end{figure}

Consider the effective Yukawa coupling operators
${\mathscr{O}}_{\alpha \beta}^\ell$ which generate the entries in the
charged lepton mass matrix via the mass terms
\begin{equation}
\mathscr{L}_Y^\ell =
\overline{L_\alpha}H_2{\mathscr{O}}^\ell_{\alpha\beta}E_\beta + {\rm h.c.},
\end{equation}
where $\alpha,\beta=e,\mu,\tau$. (See \fig~\ref{fig:effective}.)
We will denote the total number of times
that the fields $\phi_1,\phi_2,\ldots,\phi_6$ appear in the operator
${\mathscr{O}}^\ell_{\alpha \beta}$
by $n_1$ and the total number of times that their complex conjugates
$\phi_1^\dagger,\phi_2^\dagger,\ldots,\phi_6^\dagger$ appear in the operator
${\mathscr{O}}^\ell_{\alpha \beta}$ by $n_2$. Now, invariance under
transformation of the
discrete symmetry ${\mathcal{D}}_1$ implies that for the first column
of the Yukawa interaction matrix $({\mathscr{O}}^\ell_{\alpha\beta})$, \ie,
for $\beta=e$, it must hold that $n_1-n_2=4$. For the second and
third column of the Yukawa interaction matrix, \ie, for
$\beta=\mu,\tau$, the discrete symmetry ${\mathcal{D}}_1$ instead requires
that $n_1-n_2=1$.
In addition, we conclude from
the transformation properties of the fundamental Froggatt--Nielsen states under
the discrete symmetry ${\mathcal{D}}_6$ that the operators
${\mathscr{O}}^\ell_{\alpha \mu}$ and
${\mathscr{O}}^\ell_{\alpha \tau}$, where $\alpha=e,\mu,\tau$, can
neither involve the field $\phi_9$ nor the field $\phi_{10}$. This is, however,
not true for the operators ${\mathscr{O}}^\ell_{\alpha e}$ ($\alpha =
e,\mu,\tau$) in the first column of the effective Yukawa coupling matrix.

\subsection{The first row and column of the charged lepton mass matrix}
\label{sub:first}

Invariance under transformations of the U(1) symmetries requires the
U(1) charges of the entries ${\mathscr{O}}^\ell_{e\mu}$ and
${\mathscr{O}}^\ell_{e\tau}$ in the first row of the effective Yukawa
coupling matrix $({\mathscr{O}}^\ell_{\alpha\beta})$ to be
$(1,-1,0)$. Since the 
fields $\phi_9$ and $\phi_{10}$ cannot be involved in the generation of the
$e$-$\mu$- and $e$-$\tau$-elements of the charged lepton mass matrix, the U(1)
charge assignment immediately implies that any mass operator giving rise to
these $e$-$\mu$- and $e$-$\tau$-elements must involve the term
$\sim \phi_{1,2}\theta^2/(M_1)^3$. Next, the symmetries ${\mathcal{D}}_5$ and
${\mathcal{D}}_6$ yield to leading order for the operators
${\mathscr{O}}^\ell_{e\mu}$ and ${\mathscr{O}}^\ell_{e\tau}$ the two 
possible terms $\sim\phi_1\phi_1'\theta^2/(M_1)^4$ and
$\sim\phi_2\phi_2'\theta^2/(M_1)^4$. In conjuction with the requirement
$n_1-n_2=1$, the symmetry ${\mathcal{D}}_6$ implies that any
further operators contributing to the operator
${\mathscr{O}}^\ell_{e\mu}$ or ${\mathscr{O}}^\ell_{e\tau}$ must have
at least two powers of mass dimension more than the terms
$\phi_1\phi_1'\theta^2/(M_1)^4$ and
$\phi_2\phi_2'\theta^2/(M_1)^4$. We will therefore neglect these additional
operators.

{}From the transformation properties of the right-handed electron $E_e$ and the
fundamental Froggatt--Nielsen states under transformations of the
discrete symmetries ${\mathcal{D}}_1$ and ${\mathcal{D}}_6$, we conclude that
the operators ${\mathscr{O}}^\ell_{e e},{\mathscr{O}}^\ell_{\mu e}$, and
${\mathscr{O}}^\ell_{\tau e}$ in the first column of the effective
Yukawa coupling matrix
must involve at least a four-fold product of fields taken from the set
$\{\phi_1,\phi_2,\phi_3,\phi_4,\phi_5,\phi_6\}$ times a field taken from the
set $\{\phi_9,\phi_{10}\}$. Possible lowest-dimensional contributions to
the operator ${\mathscr{O}}^\ell_{ee}$, which are consistent with the
symmetries of our model, are, \eg, given by $\sim \phi_{10}
\left[(\phi_3)^4+(\phi_4)^4\right]/(M_1)^5$ and $\sim \phi_{10}
(\phi_3)^2 (\phi_4)^2 /(M_1)^5$. For brevity, we will take the operator
\begin{equation}\label{eq:Oee}
{\mathscr{O}}^\ell_{ee} =
\frac{\phi_{10}}{(M_1)^5}\left[(\phi_3)^4+(\phi_4)^4\right]
\end{equation}
as a representative of these contributions.
The remaining operators ${\mathscr{O}}^\ell_{\mu e}$ and
${\mathscr{O}}^\ell_{\tau e}$ have a mass dimension that is greater than
or equal to the mass dimension of the terms in
\eq~(\ref{eq:Oee}). However, the effects of these terms on the
leptonic mixing
angles will turn out to be negligible in comparison with the
contributions coming from other
entries of the charged lepton mass matrix.

In total, the first row of the effective Yukawa coupling matrix of the
charged leptons, which is consistent with all of the discrete
symmetries, is to leading order
\begin{subequations}\label{eq:Oealpha}
\begin{equation}
\label{eq:Oea}
 ({\mathscr{O}}^\ell_{e\alpha})=
 \left( \begin{matrix} A_1\left[(\phi_3)^4+(\phi_4)^4\right] \quad & \; 
 B_1\left[\phi_1\phi_1'-\phi_2\phi_2'\right] \quad & \;
 B_1\left[\phi_1\phi_1'+\phi_2\phi_2'\right] \end{matrix} \right).
\end{equation}
Here the dimensionful coefficients $A_1$ and $B_1$ are given by
\begin{eqnarray}
 A_1&=&Y^\ell_a\frac{\phi_{10}}{(M_1)^5}, \label{eq:A1}\\
 B_1&=&Y^\ell_b\frac{\theta^2}{(M_1)^4}, \label{eq:B1}
\end{eqnarray}
\end{subequations}
where the quantities $Y^\ell_a$ and $Y^\ell_b$ are arbitrary order
unity coefficients and $M_1$ is the high mass scale of the
intermediate Froggatt--Nielsen states. 
Note that at the level of the fundamental theory, the permutation
symmetry $\mathcal{D}_4$, which interchanges the second and third
generations of the leptons, is also propagated to the heavy
Froggatt--Nielsen states. This establishes a degeneracy of the
associated Yukawa couplings and the explicit masses of these
states, which is then translated into a degeneracy of the
corresponding effective Yukawa couplings of the low-energy theory.

\subsection{The 2-3-submatrix of the charged lepton mass matrix}
\label{sub:2-3}

In the 2-3-submatrix of the charged lepton mass matrix, the U(1) charges of the
operators ${\mathscr{O}}^\ell_{\alpha\beta}$ ($\alpha,\beta=\mu,\tau$) must be
$(0,0,0)$. The lowest dimensional
operators which fulfill this condition as well as the constraint 
$n_1-n_2=1$ are proportional to $\phi_{3,4}/M_1$ or 
$\phi_{5,6}\theta/(M_1)^2$. Furthermore,
invariance under transformation of the discrete
symmetries ${\mathcal{D}}_5$ and ${\mathcal{D}}_6$ implies that the
lowest dimensional operators ${\mathscr{O}}_{\alpha \beta}^\ell$ in
the 2-3-submatrix with $n_1-n_2=1$ are of the types
\begin{equation}
 \frac{\phi_3'\phi_3}{(M_1)^2},\quad
 \frac{\phi_4'\phi_4}{(M_1)^2},\quad
 \frac{\phi_5'\phi_5\theta}{(M_1)^3},\quad
 \frac{\phi_6'\phi_6\theta}{(M_1)^3}.
\end{equation}
Thus, the most general 2-3-submatrix of the matrix
$({\mathscr{O}}^\ell_{\alpha\beta})$, which involves only these
combinations and is invariant under transformations of the remaining
discrete symmetries, is found to be
\begin{subequations}\label{eq:23subblock}
\begin{equation}\label{eq:23subblocka}
\left(
\begin{matrix}
C (\phi_3'\phi_3-\phi_4'\phi_4) + D (\phi_5'\phi_5-\phi_6'\phi_6) & 0\\
0 & C (\phi_3'\phi_3+\phi_4'\phi_4) + D(\phi_5'\phi_5+\phi_6'\phi_6)
\end{matrix}
\right).
\end{equation}
Here the dimensionful coefficients $C$ and $D$ are given by
\begin{eqnarray}
 C &=&Y^\ell_c\frac{1}{(M_1)^2}, \label{eq:C}\\
 D &=&Y^\ell_d\frac{\theta}{(M_1)^3}, \label{eq:D}
\end{eqnarray}
\end{subequations}
where the quantities $Y^\ell_c$ and $Y^\ell_d$ are arbitrary order
unity coefficients.
Actually, contributions to the next-leading operators
${\mathscr{O}}^\ell_{\mu\tau}$ and ${\mathscr{O}}^\ell_{\tau\mu}$ are,
\eg, given by
\begin{subequations}\label{eq:mutau}
\begin{eqnarray}
\mbox{$\mu$-$\tau$}&:& \quad
\frac{1}{(M_1)^4}
(a_1\phi_7\phi_8^\dagger+a_2\phi_8\phi_7^\dagger)(\phi_3+\phi_4)\phi_3'
,\\
\mbox{$\tau$-$\mu$}&:& \quad
\frac{1}{(M_1)^4}(a_2\phi_7\phi_8^\dagger+a_1\phi_8\phi_7^\dagger)
 (\phi_3-\phi_4)\phi_3',
\end{eqnarray}
\end{subequations}
where $a_1$ and $a_2$ are some complex-valued constants. Note that the terms in
\eqs~(\ref{eq:mutau}) carry only one unit of mass dimension more
than, \eg, the term $\phi_5'\phi_5\theta/(M_1)^3$ in
\eq~(\ref{eq:23subblocka}). When diagonalizing the mass
matrix, however, it will turn out that the associated corrections to the
leptonic mixing parameters and the lepton masses are in fact negligible.

\subsection{The charged lepton mass matrix}

Combining the results of \Secs\ref{sub:first} and \ref{sub:2-3}, the
leading order effective Yukawa coupling matrix of the charged leptons
is
\begin{equation}
{\tiny
(\mathscr{O}^\ell_{\alpha\beta}) = \left( \begin{matrix}
 A_1 \left[(\phi_3)^4+(\phi_4)^4\right] &
 B_1 \left[\phi_1\phi_1'-\phi_2\phi_2'\right] &
 B_1 \left[\phi_1\phi_1'+\phi_2\phi_2'\right] \\
 0 & C (\phi_3'\phi_3-\phi_4'\phi_4) + D (\phi_5'\phi_5-\phi_6'\phi_6) & 0 \\
 0 & 0 & C (\phi_3'\phi_3+\phi_4'\phi_4) + D (\phi_5'\phi_5+\phi_6'\phi_6)
 \end{matrix} \right),
}
\end{equation}
where the dimensionful couplings $A_1$, $B_1$, $C$, and $D$ are given
in \eqs~(\ref{eq:A1}), (\ref{eq:B1}), (\ref{eq:C}), and (\ref{eq:D}),
respectively.
Inserting the VEVs in \eqs~(\ref{eq:VEVsV1}) and (\ref{eq:VEVsV3}) into the
corresponding operators of the matrix
$(\mathscr{O}^\ell_{\alpha\beta})$, we observe that, due to the vacuum
alignment mechanism of the SM singlet scalar fields, in some of the entries of
the matrix $((M_\ell)_{\alpha\beta}$), the spontaneously generated
effective mass terms of a given order exactly cancel, whereas in other
entries they do not. Furthermore, the vacuum alignment mechanism
correlates these cancellations in the different entries of the matrix
$((M_\ell)_{\alpha\beta})$ in such a way that after SSB the charged
lepton mass matrix $M_\ell$ can be of the two possible asymmetric forms
\begin{equation}
M_{\ell}\simeq m_\tau^{\rm exp}
\left(
\begin{matrix}
 \epsilon^3 & \epsilon^2 & \epsilon^4\\
  \epsilon^3 & \epsilon &\epsilon^2\\
 \epsilon^3 & \epsilon^2 & 1
\end{matrix}
\right)
\end{equation}
and
\begin{equation}
M_{\ell}\simeq m_\tau^{\rm exp}
\left(
\begin{matrix}
 \epsilon^3 & \epsilon^4 & \epsilon^2\\
 \epsilon^3 & 1 &\epsilon^2\\
 \epsilon^3 & \epsilon^2 & \epsilon
\end{matrix}
\right),
\end{equation}
where we have also introduced the appropriate orders of magnitude for the
matrix elements $(M_{\ell})_{\mu e}$ and $(M_{\ell})_{\tau e}$ as well as for
the ``phenomenological'' (in contrast to exact) texture zeros
arising in \eqs~(\ref{eq:Oea}) and (\ref{eq:23subblocka}). Here
$m_\tau^{\rm exp}$ is the (experimental) tau mass.
Note that a permutation of the second and third generations
$L_\mu\leftrightarrow L_\tau,\:E_\mu\leftrightarrow E_\tau$
leads from one solution to another. Let us consider the first
one for our remaining discussion.

\section{Yukawa interactions of the neutrinos}
\label{sec:Yukawa_n}

Consider the effective Yukawa coupling operators
${\mathscr{O}}_{\alpha\beta}^\nu$ which generate the entries in the
neutrino mass matrix $M_\nu$ via the mass terms
\begin{equation}
\mathscr{L}_Y^\nu =
\overline{L^c_{\alpha}}\frac{H_1^2}{M_2}{\mathscr{O}}_{\alpha\beta}^\nu
L_{\beta} + {\rm h.c.},
\end{equation}
where $\alpha,\beta=e,\mu,\tau$ and $M_2$ is the relevant high
mass scale which is responsible for the smallness of the
neutrino masses in comparison with the charged lepton masses.
Since the SM singlet neutrinos as well as the Higgs doublets are
${\mathcal{D}}_1$-singlets, the presence of the fields
$\phi_1,\phi_2,\ldots,\phi_6$ (which transform non-trivially under the
discrete symmetry
${\mathcal{D}}_1$) in the operators ${\mathscr{O}}^\nu_{\alpha\beta}$ is
forbidden. Hence, the only scalar fields that can be involved in the
leading order operators ${\mathscr{O}}^\nu_{\alpha\beta}$ are
$\phi_7,\phi_8,\phi_9,\phi_{10}$, and $\theta$.

\subsection{Effective Yukawa interactions of the neutrinos}

The operators
${\mathscr{O}}^\nu_{\mu\mu}, {\mathscr{O}}^\nu_{\mu\tau},
{\mathscr{O}}^\nu_{\tau\mu}$, and ${\mathscr{O}}^\nu_{\tau\tau}$ must
have the U(1) charge structure $(0,-2,0)$. An
example of the lowest dimensional operators which achieve this is
\begin{equation}\label{eq:nu23subblock}
 \sim\frac{\phi_9\phi_{10}\theta}{(M_1)^5}
 \left[(\phi_7^\dagger)^2+(\phi_8^\dagger)^2\right].
\end{equation}
Since the U(1) charge structure of the entry ${\mathscr{O}}^\nu_{ee}$ is
$(-2,0,0)$, the operator ${\mathscr{O}}^\nu_{ee}$ is to leading order
\begin{equation}\label{eq:nuOee}
 {\mathscr{O}}^\nu_{ee}= Y_a^\nu 
 \frac{\phi_9\phi_{10}\theta}{(M_1)^3},
\end{equation}
where $Y_a^\nu$ is an arbitrary order unity coefficient. Note that the
operators of the type in \eq~(\ref{eq:nu23subblock}) carry two units
of mass dimension more than the operator ${\mathscr{O}}^\nu_{ee}$.

The operators ${\mathscr{O}}^\nu_{e\mu}$ and
${\mathscr{O}}^\nu_{e\tau}$ (as well as ${\mathscr{O}}^\nu_{\mu e}$ and
${\mathscr{O}}^\nu_{\tau e}$) must have the U(1) charge structure $(-1,-1,0)$.
Therefore, the lowest dimensional terms which contribute to these
operators are $\sim \phi_7/M_1$ and $\sim\phi_8/M_1$. The symmetries
${\mathcal{D}}_2, {\mathcal{D}}_3$, and ${\mathcal{D}}_4$ then yield
to leading order ${\mathscr{O}}_{e\mu}^\nu = Y_b^\nu \phi_7/M_1$ and
${\mathscr{O}}_{e\tau}^\nu = Y_b^\nu\phi_8/M_1$, where $Y_b^\nu$ is an
arbitrary order unity coefficient. Note that the operator
${\mathscr{O}}^\nu_{ee}$ in \eq~(\ref{eq:nuOee}) carries two units of mass
dimension more than the operators
${\mathscr{O}}_{e\mu}^\nu\sim \phi_7/M_1$ and
${\mathscr{O}}_{e\tau}^\nu\sim \phi_8/M_1$. We will therefore not in
detail consider the structure of the highly suppressed terms that appear in
the 2-3-submatrix of the neutrino mass matrix as the operator given in
\eq~(\ref{eq:nu23subblock}).

\begin{figure}[ht!]
\begin{center}
\includegraphics*[bb = 130 590 490 710]{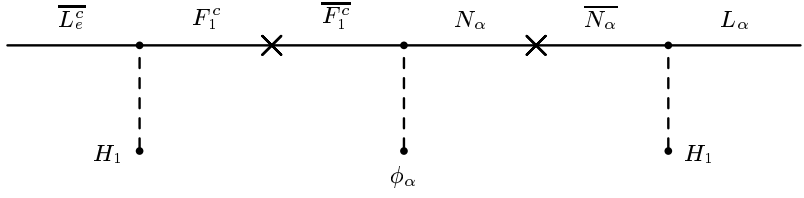}
\end{center}
\caption{The dimension six operator for $\alpha = \mu,\tau$ and
$\phi_\mu \equiv \phi_7$, $\phi_\tau \equiv \phi_8$ generating the
$e$-$\mu$- and $e$-$\tau$-elements in the effective neutrino mass matrix.}
\label{fig:dimsix}
\end{figure}

\begin{figure}[ht!]
\begin{center}
\includegraphics*[bb = 100 590 600 690, width=13.75cm]{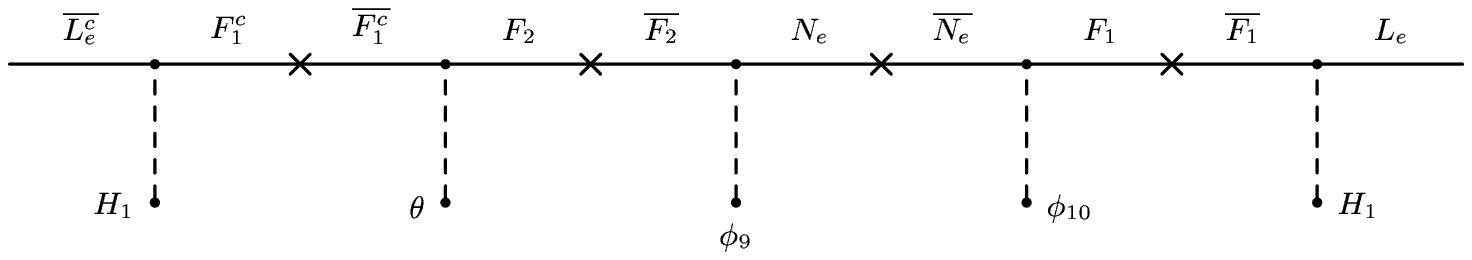}
\includegraphics*[bb = 100 590 600 690, width=13.75cm]{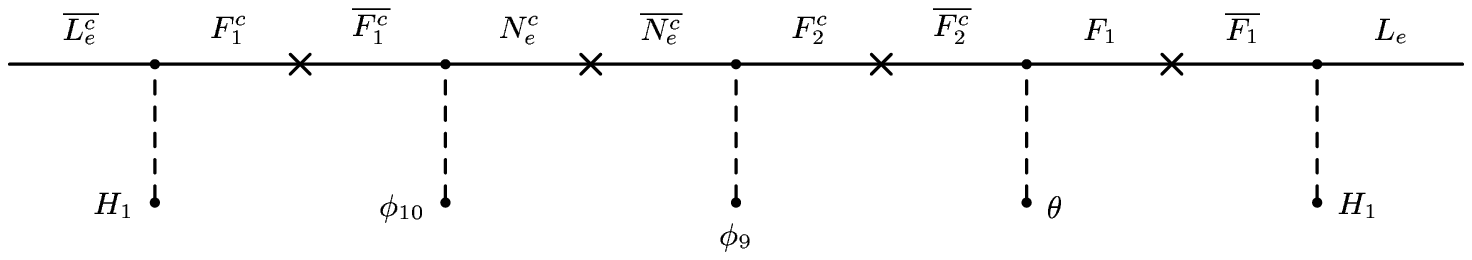}
\end{center}
\caption{The dimension eight operators generating the
$e$-$e$-element in the effective neutrino mass
matrix.}
\label{fig:dimeight}
\end{figure}

In total, the most general effective Yukawa coupling matrix of the neutrinos
$({\mathscr{O}}_{\alpha\beta}^\nu)$
that is consistent with the symmetries of our model is to leading order
given by
\begin{equation}\label{eq:Onu}
 ({\mathscr{O}}^\nu_{\alpha\beta})=
 \left(
 \begin{matrix}
  A_2 & B_2 & B_3\\
  B_2 & 0 & 0\\
  B_3 & 0 & 0
 \end{matrix}
 \right). 
\end{equation}
Here the dimensionful coefficients $A_2$, $B_2$, and $B_3$ are given by
\begin{subequations}\label{eq:A2B3B4}
\begin{eqnarray}
 A_2&=&Y_a^\nu\frac{\phi_9\phi_{10}\theta}{(M_1)^3},\\
 B_2&=&Y_b^\nu\frac{\phi_7}{M_1},\\
 B_3&=&Y_b^\nu\frac{\phi_8}{M_1},
\end{eqnarray}
\end{subequations}
where the quantity $Y^\nu_b$ is an order unity coefficient. The
leading order tree level realizations of the higher-dimensional
operators, which generate the neutrino masses, are shown in
\figs~\ref{fig:dimsix} and \ref{fig:dimeight}.
Note that the coefficients $B_2$ and $B_3$ contain the same Yukawa coupling
constant $Y^\nu_b$. Furthermore, we point out that
the texture zeros in the 2-3-submatrix of the effective neutrino Yukawa matrix
should be understood as ``phenomenological'' ones, since they actually
represent highly suppressed operators carrying two units of mass
dimension more than the entry ${\mathscr{O}}^\nu_{ee}$.

\subsection{The neutrino mass matrix}

Inserting the VEVs in \eq~(\ref{eq:VEVsV78}) into the effective operators in
\eqs~(\ref{eq:A2B3B4}), we obtain from \eq~(\ref{eq:Onu}) the
effective neutrino mass matrix (to leading order)
\begin{equation}
M_\nu = \frac{\langle H_1\rangle^2}{M_2}
\left(\begin{matrix}
  Y_a^\nu\epsilon^3 & Y_b^\nu \epsilon & \pm Y_b^\nu\epsilon\\
  Y_b^\nu\epsilon  & \epsilon^5 & \epsilon^5 \\
  \pm Y_b^\nu\epsilon  & \epsilon^5 & \epsilon^5 \\
\end{matrix} \right),
\label{eq:Mnu}
\end{equation}
where we have introduced the actual orders of magnitude of the higher-order
corrections to the texture zeros in the 2-3-submatrix of the
matrix in \eq~(\ref{eq:Onu}).
Note that after SSB the symmetries determine the
$e$-$\mu$- and $e$-$\tau$-elements to be exactly degenerate (up to a
sign), giving rise to an atmospheric mixing angle which is close to
maximal (higher-order corrections to exact maximal atmospheric mixing mainly
come from the $\mu$-$\tau$- and $\tau$-$\mu$-elements of the charged
lepton mass matrix). Introducing an ``absolute'' neutrino mass scale
$m_\nu$ and choosing $Y^\nu_a/Y^\nu_b \simeq 1$, we can write the
neutrino mass matrix in \eq~(\ref{eq:Mnu}) as
\begin{equation}
M_\nu \simeq m_\nu \left( \begin{matrix} \epsilon^2 & 1 & -1 \\ 1 & 
  \epsilon^4 & \epsilon^4 \\ -1 & \epsilon^4 & \epsilon^4 \end{matrix}
\right),
\label{eq:Mnu_app}
\end{equation}
where
\begin{equation}
m_\nu = \frac{\langle H_1 \rangle^2}{M_2} Y^\nu_b \epsilon.
\end{equation}
Note that we have chosen the minus signs for the $e$-$\tau$- and
$\tau$-$e$-elements due to our freedom of absorbing the corresponding
phase into the order unity coefficients in the charged lepton
sector. Furthermore, it is important to keep in mind that the entries
``1'' and ``$-1$'' of the matrix in \eq~(\ref{eq:Mnu_app}) indeed
denote matrix elements, which are degenerate to a very high accuracy,
whereas the other entries are only known up to their order unity coefficients.

\section{Lepton masses and leptonic mixings}
\label{sec:massmix}

{}From the results of the last two sections we have seen that the lepton mass
matrices are given by
$$
M_\ell \simeq m_\tau^{\rm exp} \left( \begin{matrix} \epsilon^3 &
  \epsilon^2 & \epsilon^4
  \\ \epsilon^3 & \epsilon & \epsilon^2 \\ \epsilon^3 & \epsilon^2 & 1
  \end{matrix} \right)
\quad \mbox{and} \quad
M_\nu \simeq m_\nu \left( \begin{matrix} \epsilon^2 & 1 & -1 \\ 1 &
  \epsilon^4 & \epsilon^4 \\ -1 & \epsilon^4 & \epsilon^4 \end{matrix}
\right),
$$
where again $m_\tau^{\rm exp}$ is the (experimental) tau mass, $m_\nu$ is some
``absolute'' neutrino mass scale, $\epsilon \simeq 0.1$ is the
small expansion parameter (defined in \Sec\ref{sec:scalar_pot}), and
only the order of magnitude of the matrix elements have been indicated. In the
above expression, the first matrix is the charged lepton mass matrix and
the second matrix is the neutrino mass matrix. Note that the small
expansion parameter $\epsilon$ is the same for both matrices. In order to
find the leptonic mixing angles and the lepton masses, we have to perform
diagonalizations of the above two displayed mass matrices.
Let us begin with the diagonalization of the charged lepton mass
matrix. Since this matrix is not symmetric and we want to extract the
relevant mass and mixing parameters, we have, in fact, to
diagonalize the matrix product $M_\ell M_\ell^\dagger$, which is a
symmetric matrix. Biunitary
diagonalization of the matrix $M_\ell$ implies that $U_\ell^\dagger M_\ell
V_\ell = \mathcal{M}_\ell$, where $\mathcal{M}_\ell = {\rm diag
  \,}(m_e,m_\mu,m_\tau)$ is the diagonalized charged lepton mass
matrix containing the masses of the charged leptons (\ie, the
eigenvalues of the matrix $\sqrt{M_\ell M_\ell^\dagger}$) 
and $U_\ell$ and $V_\ell$ are
two unitary mixing matrices. Thus, we obtain $\mathcal{M}_\ell
\mathcal{M}_\ell^\dagger = U_\ell^\dagger M_\ell M_\ell^\dagger
U_\ell$, where $U_\ell$ will be the charged lepton mixing matrix. Note
that the mixing matrix $V_\ell$ does not appear in the diagonalization
of the matrix product $M_\ell M_\ell^\dagger$. Actually, since the matrix
$M_\ell$ is a real matrix, we have $\mathcal{M}_\ell
\mathcal{M}_\ell^\dagger = \mathcal{M}_\ell \mathcal{M}_\ell^T =
U_\ell^T M_\ell M_\ell^T U_\ell$. Next, straight-forward calculations
(using \App\ref{app:standard}) yield
\begin{equation}
\mathcal{M}_\ell \mathcal{M}_\ell^\dagger = (m_\tau^{\rm exp})^2 \; {\rm diag
  \,}(\epsilon^6 - 2 \epsilon^7 + \mathcal{O}(\epsilon^9), \epsilon^2
  + \epsilon^4 + \mathcal{O}(\epsilon^5), 1 + 2 \epsilon^4 +
\mathcal{O}(\epsilon^5))
\end{equation}
and
\begin{equation}
U_\ell \equiv ((U_\ell)_{\alpha a}) \simeq \left( \begin{matrix}
  -0.995 & -0.0997 & 
  -0.000212 \\ 0.0997 & -0.995 & -0.0111 \\ -0.000897 &
  0.0111 & -1.000 \end{matrix} \right)
\label{eq:Ul}
\end{equation}
such that $U_\ell^\dagger U_\ell \simeq U_\ell U_\ell^\dagger \simeq
1_3$, where $\alpha = e,\mu,\tau$ and $a = 1,2,3$. Note that the charged
lepton mixing
matrix $U_\ell$ is independent of the absolute charged lepton mass
scale $m_\tau^{\rm exp}$, \ie, the tau mass. This means that the
charged lepton (and, of course, the leptonic) mixing angles will not
depend on the tau mass. The charged lepton masses are given as the
square roots of (the absolute values of) the eigenvalues of the matrix
product $M_\ell M_\ell^\dagger$, \ie, $m_e = 
m_\tau^{\rm exp} \epsilon^3 \left( 1 - \epsilon + \mathcal{O}(\epsilon^2)
\right)$, $m_\mu = m_\tau^{\rm exp} \epsilon \left( 1 + \tfrac{1}{2} \epsilon^2
+ \mathcal{O}(\epsilon^3) \right)$, and $m_\tau = m_\tau^{\rm exp} \left( 1
+ \epsilon^4 + \mathcal{O}(\epsilon^5) \right)$.
Thus, the order-of-magnitude relations for the charged lepton
masses are given by
\begin{equation}
m_e/m_\mu \simeq \epsilon^2 \simeq 10^{-2} \quad \mbox{and} \quad
m_\mu/m_\tau \simeq \epsilon \simeq 10^{-1},
\end{equation}
which approximately fit the experimentally observed values, \ie,
$(m_e/m_\mu)_{\rm exp} \simeq 4.8 \cdot 10^{-3}$ and
$(m_\mu/m_\tau)_{\rm exp} \simeq 5.9 \cdot 10^{-2}$ \cite{groo00}.
Furthermore, the charged lepton mixing angles (in the ``standard''
parameterization) are found to be (using \App\ref{app:standard})
\cite{Ohlsson:1999xb,Ohlsson:2001vp,Ohlsson:2002na}
\begin{eqnarray}
\theta_{12}^\ell &\equiv& \arctan \left| \frac{(U_\ell)_{e
  2}}{(U_\ell)_{e 1}} \right| 
  = \epsilon - \frac{1}{3} \epsilon^3 + \mathcal{O}(\epsilon^4),
  \label{eq:12l} \\
\theta_{13}^\ell &\equiv& \arcsin |(U_\ell)_{e 3}| = 2 \epsilon^2 +
\epsilon^5 + \mathcal{O}(\epsilon^6), \label{eq:13l} \\
\theta_{23}^\ell &\equiv& \arctan \left| \frac{(U_\ell)_{\mu
    3}}{(U_\ell)_{\tau 3}} \right| = \epsilon^2 + \epsilon^3 +
\mathcal{O}(\epsilon^4). \label{eq:23l}
\end{eqnarray}
Inserting $\epsilon \simeq 0.1$ into \eqs~(\ref{eq:12l}) -
(\ref{eq:23l}), we obtain $\theta_{12}^\ell \simeq 5.72^\circ
\approx 6^\circ$,
$\theta_{13}^\ell \simeq 0.0122^\circ \approx 0.01^\circ$, and
$\theta_{23}^\ell \simeq 0.637^\circ \approx 0.6^\circ$, \ie, the
mixing angles in the charged lepton sector are all small. The mixing
angle $\theta^\ell_{12}$ is the only one that is not negligible. Thus,
it will finally yield a non-zero contribution to the leptonic mixing
angles \cite{Ohlsson:2002na}.

Next, we want to diagonalize the neutrino mass matrix $M_\nu$. The
diagonalization procedure of the matrix $M_\nu$ is easier than the one of the
matrix $M_\ell$, since the matrix $M_\nu$ is a symmetric matrix.
Diagonalization of the matrix $M_\nu$ directly implies that $U_\nu^T M_\nu
U_\nu = \mathcal{M}_\nu$, where $\mathcal{M}_\nu = {\rm diag
  \,}(m_{\nu_1}, m_{\nu_2}, m_{\nu_3})$ is the diagonalized neutrino
mass matrix containing the masses of the neutrino mass eigenstates
(\ie, the eigenvalues of the matrix $M_\nu$)
and $U_\nu$ is the unitary neutrino mixing matrix. Straight-forward
calculations (using \App\ref{app:standard}) yield
\begin{eqnarray}
\mathcal{M}_\nu &=& m_\nu \; {\rm diag \,}\left(\frac{1}{2}
\left(\epsilon^2 + \sqrt{8 + \epsilon^4}\right), \frac{1}{2}
\left(\epsilon^2 - \sqrt{8 + \epsilon^4}\right), 2 \epsilon^4\right)
\nonumber\\
&=& m_\nu \; {\rm diag \,} \left(\sqrt{2} + \frac{1}{2} \epsilon^2
+ \frac{1}{8 \sqrt{2}} \epsilon^4 + \mathcal{O}(\epsilon^8), -
\sqrt{2} + \frac{1}{2} \epsilon^2 - \frac{1}{8 \sqrt{2}} \epsilon^4 +
\mathcal{O}(\epsilon^8), 2 \epsilon^4 \right) \nonumber\\
\end{eqnarray}
and
\begin{equation}
U_\nu \equiv ((U_\nu)_{\alpha a}) \simeq \left( \begin{matrix} 0.708 &
  0.706 & 0 \\ 0.499
  & -0.501 & \frac{1}{\sqrt{2}} \\ -0.499 & 0.501 &
  \frac{1}{\sqrt{2}} \end{matrix} \right)
\label{eq:Un}
\end{equation}
such that $U_\nu^\dagger U_\nu \simeq U_\nu U_\nu^\dagger \simeq 1_3$,
where $\alpha = e,\mu,\tau$ and $a = 1,2,3$. Note that the neutrino mixing
matrix $U_\nu$ is independent of the absolute neutrino mass
scale $m_\nu$. This means that the neutrino (and, of course, the leptonic)
mixing angles will not depend on this scale. The (physical) neutrino
masses are given as (the absolute values of) the eigenvalues
of the neutrino mass matrix $M_\nu$, \ie, $m_1 \equiv |m_{\nu_1}| \simeq
\sqrt{2} m_\nu$, $m_2 \equiv |m_{\nu_2}| \simeq \sqrt{2} m_\nu$, and $m_3
\equiv |m_{\nu_3}| = 2 \epsilon^4 m_\nu \approx 0$, which means that
we have an inverted neutrino mass hierarchy spectrum (\ie, $m_3 \ll
m_1 \simeq m_2 \quad \Rightarrow \quad 0 \simeq |\Delta m_{21}^2| \ll
|\Delta m_{32}^2| \simeq |\Delta m_{31}^2|$).
Naturally, the neutrino mass 
squared differences $\Delta m_{ab}^2 \equiv m_{\nu_a}^2 - m_{\nu_b}^2
= m_a^2 - m_b^2$,
where $m_{\nu_a}$ is the mass and $m_a$ is the physical mass of the
$a$th neutrino mass eigenstate, respectively, are given as follows:
\begin{eqnarray}
\Delta m_{21}^2 &=& m_\nu^2 \left( - 2 \sqrt{2} \epsilon^2 -
\frac{1}{4 \sqrt{2}} \epsilon^6 + \mathcal{O}(\epsilon^{10}) \right),
\label{eq:dm21} \\
\Delta m_{32}^2 &=& m_\nu^2 \left( - 2 + \sqrt{2} \epsilon^2 -
\frac{1}{2} \epsilon^4 + \mathcal{O}(\epsilon^6) \right),
\label{eq:dm32} \\
\Delta m_{31}^2 &=& m_\nu^2 \left( - 2 - \sqrt{2} \epsilon^2 -
\frac{1}{2} \epsilon^4 + \mathcal{O}(\epsilon^6)
\right). \label{eq:dm31}
\end{eqnarray}
Note that the leading order terms in \eqs~(\ref{eq:dm21}) -
(\ref{eq:dm31}) are all negative, which is natural, since we have an
inverted mass hierarchy spectrum for the neutrinos. Hence, the solar
and atmospheric neutrino mass squared differences are given by $\Delta
m_\odot^2 \equiv |\Delta m_{21}^2| \simeq 2 \sqrt{2} \epsilon^2
m_\nu^2$ and $\Delta m_{\rm atm}^2 \equiv |\Delta m_{32}^2| \simeq
|\Delta m_{31}^2| \simeq 2 m_\nu^2$, respectively. The
``experimental'' values of these quantities are $\Delta m_\odot^2
\simeq 5.0 \cdot 10^{-5} \, {\rm eV}^2$ \cite{Bahcall:2002hv} and
$\Delta m_{\rm atm}^2 \simeq 2.5 \cdot 10^{-3} \,
{\rm eV}^2$ \cite{Toshito:2001dk}.\footnote{@ 99.73\% C.L.: $2.3 \cdot
  10^{-5} \, {\rm eV}^2 \lesssim \Delta m_\odot^2 \lesssim 3.7 \cdot
  10^{-4} \, {\rm eV}^2$ \cite{Bahcall:2002hv}; best-fit:
  $\Delta m_\odot^2 \simeq 5.0 \cdot 10^{-5} \, {\rm eV}^2$
  \cite{Bahcall:2002hv}}\footnote{@ 90\% C.L.: $1.6 \cdot 10^{-3} \, {\rm eV}^2
  \lesssim \Delta m_{\rm atm}^2 \lesssim 3.9 \cdot 10^{-3} \, {\rm
    eV}^2$ \cite{Shiozawa:2002}; best-fit: $\Delta m_{\rm atm}^2
  \simeq 2.5 \cdot 10^{-3}\, {\rm eV}^2$ \cite{Toshito:2001dk}}
Thus, we extract that $m_\nu \simeq
0.04 \, {\rm eV}$ and $\epsilon \simeq 0.1$, which is consistent and
agrees perfectly with our used value for the small expansion parameter
$\epsilon$. In other words, using $\epsilon \simeq 0.1$ and $m_\nu
\simeq 0.04 \, {\rm eV}$ for the absolute neutrino mass scale (Nobody
knows where this value comes from!), we obtain the presently preferred
values for the solar and atmospheric neutrino mass squared differences.
Furthermore, we have that $m_1 \simeq m_2 \simeq 0.05 \,
{\rm eV}$ and $m_3 \simeq 1 \cdot 10^{-5} \, {\rm eV} \approx 0$. 
Similarly, as for the charged leptons, the neutrino mixing angles (in
the ``standard'' parameterization) are found to be (using
\App\ref{app:standard}) \cite{Ohlsson:1999xb,Ohlsson:2001vp,Ohlsson:2002na}
\begin{eqnarray}
\theta_{12}^\nu &\equiv& \arctan \left|
\frac{(U_\nu)_{e2}}{(U_\nu)_{e1}} \right| =
\frac{\pi}{4} + \frac{1}{4 \sqrt{2}} \epsilon^2 - \frac{1}{96
  \sqrt{2}} \epsilon^6 + \mathcal{O}(\epsilon^{10}), \label{eq:12n} \\
\theta_{13}^\nu &\equiv& \arcsin |(U_\nu)_{e3}| = 0, \label{eq:13n} \\
\theta_{23}^\nu &\equiv& \arctan \left|
\frac{(U_\nu)_{\mu 3}}{(U_\nu)_{\tau 3}} \right| =
\frac{\pi}{4}. \label{eq:23n}
\end{eqnarray}
Inserting $\epsilon \simeq 0.1$ into \eqs~(\ref{eq:12n}) -
(\ref{eq:23n}), we obtain $\theta_{12}^\nu \simeq 45.1^\circ
\approx 45^\circ$, $\theta_{13}^\nu = 0$, and $\theta_{23}^\nu =
45^\circ$, which means that we have (nearly) bimaximal mixing in the
neutrino sector. Note that the values for $\theta_{13}^\nu$ and
$\theta_{23}^\nu$ are exact, since they are only determined from the
matrix elements of the third column of the matrix $U_\nu$ in \eq~(\ref{eq:Un}).

\section{The leptonic mixing angles}
\label{sec:leptonic}

The leptonic mixing angles (or parameters if one also includes the
$\mathcal{CP}$ violation phase $\delta$) are given by the leptonic mixing
matrix\footnote{The leptonic mixing matrix is sometimes called the
  Maki--Nakagawa--Sakata (MNS) mixing matrix \cite{Maki:1962mu}.}.
The leptonic mixing matrix $U$ is composed of the charged lepton
mixing matrix $U_\ell$ and the neutrino mixing matrix $U_\nu$ as
follows:
\begin{equation}
U \equiv U_\ell^\dagger U_\nu. \label{eq:U}
\end{equation}
The matrix $U_\ell$ ($U_\nu$) rotates the left-handed charged lepton
fields (the neutrino fields) so that the charged lepton mass matrix $M_\ell$
(the neutrino mass matrix $M_\nu$) becomes diagonal (see
\Sec\ref{sec:massmix}), \ie, it relates the flavor state and mass eigenstate
bases. Thus, one can look upon the matrix $U$ [$U_{ab} =
  (U_\ell^\dagger U_\nu)_{ab} = \sum_{\alpha = e,\mu,\tau}
  (U_\ell^\dagger)_{a\alpha} (U_\nu)_{\alpha b} = \sum_{\alpha =
    e,\mu,\tau} (U_\ell)^\ast_{\alpha a} (U_\nu)_{\alpha b}$] as the
messenger between the mass eigenstate bases of the charged leptons and
the neutrinos.
Inserting the matrices in \eqs~(\ref{eq:Ul}) and (\ref{eq:Un}) into
the definition in \eq~(\ref{eq:U}), we find that
\begin{equation}
U \equiv (U_{ab}) \simeq \left( \begin{matrix} -0.655 &
  -0.753 & 0.0699 \\ 
  -0.573 & 0.434 & -0.696 \\ 0.493 & -0.495 & -0.715
\end{matrix} \right),
\label{eq:U_app}
\end{equation}
which fulfill the condition $U^\dagger U = U U^\dagger = 1_3$ to a
very good accuracy.
Now, the leptonic mixing angles (in the ``standard'' parameterization
[see \App\ref{app:standard}])
can be read off as follows \cite{Ohlsson:1999xb,Ohlsson:2001vp,Ohlsson:2002na}:
\begin{eqnarray}
\theta_{12} &\equiv& \arctan \left| \frac{U_{12}}{U_{11}} \right|,
\label{eq:12} \\
\theta_{13} &\equiv& \arcsin |U_{13}|, \label{eq:13} \\
\theta_{23} &\equiv& \arctan \left| \frac{U_{23}}{U_{33}}
\right|. \label{eq:23}
\end{eqnarray}
Thus, inserting the appropriate matrix elements of the matrix $U$
expressed in terms of the small expansion parameter $\epsilon$
($\epsilon \ll 1$) in \eq~(\ref{eq:U}) into \eqs~(\ref{eq:12}) -
(\ref{eq:23}), we obtain
\begin{eqnarray}
\theta_{12} &=& \frac{\pi}{4} - \frac{1}{\sqrt{2}} \epsilon +
\frac{1}{4 \sqrt{2}} \epsilon^2 + \mathcal{O}(\epsilon^3), \label{eq:t12}\\
\theta_{13} &=& \frac{1}{\sqrt{2}} \epsilon - \frac{17}{12 \sqrt{2}}
\epsilon^3 + \mathcal{O}(\epsilon^4), \label{eq:t13}\\
\theta_{23} &=& \frac{\pi}{4} - \frac{5}{4} \epsilon^2 - \epsilon^3 +
\mathcal{O}(\epsilon^4). \label{eq:t23}
\end{eqnarray}
Note that the first correction
to the atmospheric (neutrino) mixing angle $\theta_{23}$ is of second
order in the small expansion parameter $\epsilon$, and it is therefore
very small, \ie, the atmospheric mixing angle stays nearly
maximal, $\theta_{23} \simeq \tfrac{\pi}{4}$. However, the first
corrections to the solar (neutrino) mixing
angle $\theta_{12}$ and the reactor (neutrino) mixing angle (the
so-called CHOOZ mixing angle) $\theta_{13}$ are both of first order in
the small expansion parameter $\epsilon$ and they are of exactly the
same size, but with opposite sign. Thus, we have the first order relation
$\theta_{12} \simeq \tfrac{\pi}{4} - \theta_{13}$.
Finally, inserting $\epsilon \simeq 0.1$ in \eqs~(\ref{eq:t12}) -
(\ref{eq:t23}), we find that the leptonic mixing angles are
$$
\theta_{12} \simeq 41.0^\circ \approx 41^\circ, \quad \theta_{13} \simeq
4.01^\circ \approx 4^\circ, \quad \mbox{and} \quad \theta_{23}
\simeq 44.2^\circ \approx 44^\circ,
$$
which means that we have bilarge leptonic mixing.
These values of the leptonic mixing angles lie
within the ranges preferred by the MSW LMA solution of the solar
neutrino problem\footnote{@ 99.73\% C.L.: $0.24 \lesssim
\tan^2 \theta_{12} \lesssim 0.89 \quad \Rightarrow \quad 26^\circ
\lesssim |\theta_{12}| \lesssim 43^\circ$
\cite{Bahcall:2002hv}; best-fit: $\tan^2 \theta_{12} 
\simeq 0.42 \quad \Rightarrow \quad |\theta_{12}| \simeq 33^\circ$
\cite{Bahcall:2002hv}}, atmospheric neutrino data\footnote{@ 90\%
  C.L.: $\sin^2 2\theta_{23} \gtrsim 0.92$ \cite{Shiozawa:2002};
  best-fit: $\sin^2 2\theta_{23} \simeq 1.00 \quad \Rightarrow \quad
  |\theta_{23}| \simeq 45.0^\circ$ \cite{Toshito:2001dk}} (nearly
maximal atmospheric mixing), and CHOOZ reactor neutrino 
data. The so-called CHOOZ upper bound is $\sin^2 \theta_{13} \lesssim
0.10$ (\ie, $|\theta_{13}| \lesssim 9.2^\circ$)
\cite{Apollonio:1998xe,Apollonio:1999ae,Bemporad:1999de}.
In particular, the obtained value for the solar mixing angle
$\theta_{12}$ implies a significant deviation from maximal solar
mixing. However, the solar mixing angle is bounded from below by
approximately $41^\circ$ and it is therefore still too close to
maximal to be in the 90\% (or 95\% or 99\%) confidence level region of
the MSW LMA solution \cite{Bahcall:2002hv}.

As we have seen the symmetries of our model generate the hierarchical
charged lepton mass spectrum as well as an essentially maximal
atmospheric mixing angle. However, these symmetries only determine the
order of magnitude of the entries $(M_\ell)_{\alpha\beta}$ in the
charged lepton mass matrix $M_\ell$. In order to decide whether the
model can naturally give the MSW LMA solution or not, we can test the
robustness of the above calculated leptonic mixing angles (and charged
lepton masses) under variation of the involved order unity coefficients.
For example, by changing the ratio of the order unity coefficients
$Y_b^\ell/Y_d^\ell$ from 1 to 2 leads to
$$
\theta_{12} \simeq 37^\circ, \quad \theta_{13} \simeq 8^\circ, \quad \mbox{and}
\quad \theta_{23} \simeq 44^\circ,
$$
where the new value for $\theta_{12}$ lies within the 90\% (and 95\% and
99\% and 99.73\%) confidence level region of the MSW LMA solution
\cite{Bahcall:2002hv} and the new value for $\theta_{13}$ is still
below the CHOOZ upper bound. Note that the new values for
$\theta_{23}$ is, in principle, the same as the old one.
At the same time, the exact values of the charged lepton masses can be
accommodated by choosing the values $Y^\ell_a = 0.5$, $Y^\ell_c = 1.8$,
and $Y^\ell_d = 1.0$ for the order unity coefficients. Hence, our
model is in prefect agreement with the MSW LMA solution and it can
reproduce the realistic charged lepton mass spectrum.

\section{Implications for neutrinoless double
  $\mbox{\boldmath{$\beta$}}$-decay, astrophysics, and cosmology}
\label{sec:impl}

Assuming massive Majorana neutrinos, we analyze the implication for
the prediction of the effective Majorana mass $|\langle m \rangle|$ in 
neutrinoless double $\beta$-decay (see, \eg,
\Refs\cite{Bilenky:1987ty,Bilenky:2001rz})
\begin{equation}
|\langle m \rangle| \equiv \left| \sum_{a=1}^3 U_{1a}^2 m_{\nu_a} \right|,
\label{eq:ndbd}
\end{equation}
where the $U_{1a}$'s are first row matrix elements of the leptonic
mixing matrix $U$ and the $m_{\nu_a}$'s are the masses of the neutrino mass
eigenstates. Inserting the expressions for the $U_{1a}$'s and the
$m_{\nu_a}$'s into \eq~(\ref{eq:ndbd}), we obtain
\begin{eqnarray}
|\langle m \rangle| &=& \frac{1}{2} m_\nu \left[ \left( \sqrt{8 +
 \epsilon^4} \cos 2 \theta_{12} + \epsilon^2 \right) \cos^2
 \theta_{13} + 4 \epsilon^4 \sin^2 \theta_{13} \right] \nonumber\\
&=& m_\nu \bigg[ \sqrt{2} \cos 2 \theta_{12} \cos^2 \theta_{13} +
 \frac{1}{2} \cos^2 \theta_{13} \epsilon^2 \nonumber\\
&& + \left( \frac{1}{8
 \sqrt{2}} \cos 2 \theta_{12} \cos^2 \theta_{13} + 2 \sin^2
 \theta_{13} \right) \epsilon^4 + \mathcal{O}(\epsilon^8) \bigg] \nonumber\\
&\simeq& m_\nu \sqrt{2} \cos 2 \theta_{12} \cos^2 \theta_{13}.
\end{eqnarray}
Note that the quantity $|\langle m \rangle|$ becomes equal to zero
when $\theta_{12} = \tfrac{\pi}{4}$, \ie, when we have an exactly
maximal solar mixing angle.\footnote{A maximal solar mixing angle
  could, \eg, imply the presence of a superlight Dirac neutrino. For a
treatment of naturally light Dirac neutrinos using the seesaw
mechanism, see, \eg, \Ref\cite{Lindner:2002}.}
Using $m_\nu \simeq 0.04 \, {\rm eV}$, $\theta_{12} \simeq 41^\circ$,
and $\theta_{13} \simeq 4^\circ$, we find that $|\langle m \rangle|
\simeq 0.007 \, {\rm eV}$, which is consistent with and below the
phenomenological upper bound $|\langle m \rangle| < 0.080 \, {\rm eV}$
\cite{Pascoli:2002xq} for an inverted hierarchical neutrino mass
spectrum. Our value for $|\langle m \rangle|$ is also below the
experimental upper bound obtained from neutrinoless double
$\beta$-decay experiments (${}^{76}{\rm Ge}$ experiments) by
the Heidelberg-Moscow collaboration [$|\langle m \rangle| < 0.35 \,
{\rm eV}$ (@ 90\% C.L.)] \cite{Klapdor-Kleingrothaus:2000dg} and by
the IGEX collaboration [$|\langle m \rangle| < (0.33 \div 1.35) \,
{\rm eV}$ (@ 90\% C.L.)] \cite{Aalseth:2000ud,Aalseth:2002rf}.
It would also be below the sensitivity of the planned neutrinoless
double $\beta$-decay experiments GENIUS, EXO, MAJORANA, and MOON,
which is $|\langle m \rangle| \sim 0.01 \, {\rm eV}$.

Furthermore, the sum of the neutrino masses
\begin{equation}
M \equiv \sum_{a=1}^3 m_a,
\label{eq:sum}
\end{equation}
where the $m_a$'s are the physical masses of the neutrino mass
eigenstates, is often used in astrophysics and cosmology. Inserting
the expressions for the $m_a$'s into \eq~(\ref{eq:sum}), we obtain
\begin{eqnarray}
M &=& m_\nu \left( \sqrt{8 + \epsilon^2} + 2 \epsilon^4 \right)
 \nonumber\\
&=& m_\nu
\left[ 2 \sqrt{2} + \left( 2 + \frac{1}{4 \sqrt{2}} \right) \epsilon^4
  - \frac{1}{128 \sqrt{2}} \epsilon^8 + \mathcal{O}(\epsilon^{10}) \right]
  \simeq 2 \sqrt{2} m_\nu.
\end{eqnarray}
Again using $m_\nu \simeq 0.04 \, {\rm eV}$, we find that $M \simeq 0.1 \,
{\rm eV}$. There exist several upper bounds for this quantity from several
different branches of astrophysics and cosmology. For example, using
the presently best available data from cosmic microwave background
radiation and large scale structure measurements, we have the upper
bound $M < (2.5 \div 3) \, {\rm eV}$ (@ 95\% C.L.)
\cite{Hannestad:2002xv,Hannestad:2002}.
Other upper bounds of this quantity come, \eg, from cosmic microwave
background radiation, galaxy clustering, and Lyman Alpha Forest
measurements $M < 4.2 \, {\rm eV}$ (@ 95\% C.L.) \cite{Wang:2001gy} and
from the 2dF Galaxy Redshift Survey $M < (1.8 \div 2.2) \, {\rm eV}$
(@ 95\% C.L.) \cite{Elgaroy:2002bi}. Thus, our value for the sum of neutrino 
masses is well below the upper bounds derived from astrophysics and cosmology.
However, it should be possible to combine cosmic microwave background
radiation data from the MAP/Planck satellite with Sloan Digital Sky Survey
measurements, which could give an upper bound of $M < 0.3 \, {\rm eV}$
\cite{Hu:1998mj}. Then, our obtained value for the sum of the neutrino
masses is rather close to this upper bound.

\section{Summary and conclusions}
\label{sec:s&c}

In summary, we have presented a model based on flavor symmetries and
higher-dimensional mass operators. This model is a modified and
extended version of the model given in \Ref\cite{Ohlsson:2002na} and
it naturally yields the mass matrix textures
$$
\left( \begin{matrix} \epsilon^3 &
  \epsilon^2 & \epsilon^4
  \\ \epsilon^3 & \epsilon & \epsilon^2 \\ \epsilon^3 & \epsilon^2 & 1
\end{matrix} \right) \quad \mbox{and} \quad
\left( \begin{matrix} \epsilon^2 & 1 & -1 \\ 1 &
  \epsilon^4 & \epsilon^4 \\ -1 & \epsilon^4 & \epsilon^4 \end{matrix} \right)
$$
for the charged leptons and the neutrinos, respectively. Note that
these textures involve the same small expansion parameter
$\epsilon$. Our old model \cite{Ohlsson:2002na} had a symmetric mass
matrix for the charged leptons, whereas the new model has an asymmetric
one. Thus, our new model predicts the realistic charged lepton mass
spectrum (with the order unity coefficients $Y^\ell_a = 0.5$,
$Y^\ell_c = 1.8$, and $Y^\ell_d = 1.0$), an inverted hierarchical
neutrino mass spectrum, a large (but not necessarily close to maximal)
solar mixing angle $\theta_{12}$, which is in excellent agreement with
the MSW LMA solution of the solar neutrino problem, a small reactor
mixing angle $\theta_{13}$, and an approximately maximal atmospheric
mixing angle $\theta_{23}$ (enforced by the flavor
symmetries). Furthermore, these textures yield the MSW LMA solution
and it follows from the model that $\theta_{12} \simeq \tfrac{\pi}{4}
- \theta_{13}$. The explicitly obtained values for the mixing angles
(assuming no $\mathcal{CP}$ violation, \ie, $\delta = 0$) are
$$
\theta_{12} \simeq 41^\circ, \quad \theta_{13} \simeq 4^\circ, \quad
\mbox{and} \quad \theta_{23} \simeq 44^\circ
$$
for the ratio of the order unity coefficients $Y^\ell_b/Y^\ell_d = 1$ and
$$
\theta_{12} \simeq 37^\circ, \quad \theta_{13} \simeq 8^\circ, \quad
\mbox{and} \quad \theta_{23} \simeq 44^\circ,
$$
for the ratio of the order unity coefficients $Y^\ell_b/Y^\ell_d = 2$.
Both these sets of the leptonic mixing angles are in very good
agreement with present experimental data. In addition, our values are
not in conflict with limits coming from neutrinoless double
$\beta$-decay, astrophysics, and cosmology.

Finally, we have also (in the Appendix) presented a scheme for how to
transform any given $3 \times 3$ unitary matrix to the ``standard''
parameterization form of the Particle Data Group.

\ack
We would like to thank Martin Freund, Walter Grimus, Steve King, and
Holger Bech Nielsen for useful discussions and comments.

This work was supported by the Swedish Foundation for International
Cooperation in Research and Higher Education (STINT) [T.O.], the
Wenner-Gren Foundations [T.O.], the Magnus Bergvall Foundation
(Magn. Bergvalls Stiftelse) [T.O.], and the ``Sonderforschungsbereich
375 f{\"u}r Astro-Teilchenphysik der Deutschen
Forschungsgemeinschaft'' [T.O. and G.S.].

\begin{appendix}

\section{Transformation of any $\mbox{\boldmath{$3 \times 3$}}$
  unitary matrix to the ``standard'' parameterization form}
\label{app:standard}

The ``standard'' parameterization form of a $3 \times 3$ unitary
matrix according to the Particle Data Group \cite{groo00} reads
\begin{equation}
U = \left( \begin{matrix} C_2 C_3 & S_3 C_2 & S_2 {\rm e}^{- {\rm i} \delta}\\
- S_3 C_1 - S_1 S_2 C_3 {\rm e}^{{\rm i} \delta} & C_1 C_3 - S_1 S_2 S_3
{\rm e}^{{\rm i} \delta} & S_1 C_2\\
S_1 S_3 - S_2 C_1 C_3 {\rm e}^{{\rm i} \delta} & - S_1 C_3 - S_2 S_3 C_1
{\rm e}^{{\rm i}\delta} & C_1 C_2 \end{matrix}\right),
\label{eq:Ustand}
\end{equation}
where $S_a \equiv \sin \theta_a$, $C_a \equiv \cos \theta_a$ (for $a =
1,2,3$), and $\delta$ is the physical $\mathcal{CP}$ violation phase. Here
$\theta_1 \equiv \theta_{23}$, $\theta_2 \equiv \theta_{13}$, and
$\theta_3 \equiv \theta_{12}$ are the Euler (mixing) angles. Note that
four of the entries of the matrix $U$ in \eq~(\ref{eq:Ustand}) are
real, \ie, the 1-1, 1-2, 2-3, and 3-3 entries. Furthermore, the matrix
$U$ in \eq~(\ref{eq:Ustand}) can be decomposed as follows:
\begin{eqnarray}
U &=& O_{23}(\theta_{23}) U_{13}(\theta_{13},\delta)
O_{12}(\theta_{12}) \nonumber\\
&=& \left(\begin{matrix} 1 & 0 & 0\\ 0 & C_{23} & S_{23}\\ 0 & - S_{23} &
C_{23} \end{matrix}\right) \left(\begin{matrix} C_{13} & 0 & S_{13}
{\rm e}^{- {\rm i} \delta}\\ 0 & 1 & 0\\ - S_{13} {\rm e}^{{\rm i}
\delta} & 0 & C_{13} \end{matrix}\right)
\left(\begin{matrix} C_{12} & S_{12} & 0\\
- S_{12} & C_{12} & 0\\ 0 & 0 & 1 \end{matrix}\right),
\label{eq:Ustand_decomp}
\end{eqnarray}
where $C_{ab} \equiv \cos \theta_{ab}$, $S_{ab} \equiv \sin
\theta_{ab}$ (for $a,b = 1,2,3$), and $O_{ab}(\theta_{ab})$ is a
rotation by an angle $\theta_{ab}$ in the $ab$-plane. If $\delta = 0$, then
$U_{13}(\theta_{13},0) = O_{13}(\theta_{13})$.

Here we will show how to transform any given $3 \times 3$ unitary matrix
to the ``standard'' parameterization form.
A $3 \times 3$ unitary matrix
\begin{equation}
U \equiv (U_{ab}) = \left( \begin{matrix} U_{11} & U_{12} &
  U_{13} \\ U_{21} & U_{22} & U_{23} \\ U_{31} & U_{32} & U_{33}
  \end{matrix} \right)
\end{equation}
where $a,b = 1,2,3$, obeys $U^\dagger U = U
U^\dagger = 1_3$ (9 unitarity conditions) and is therefore
characterized by 9 real 
parameters, since a general $3 \times 3$ complex matrix is
characterized by 18 real parameters (9 complex parameters). Out of
these 9 parameters, 3 are Euler (mixing) angles and 
6 are phase factors (phases). Not all of the 6 phases enter into
expressions for physically measurable quantities. Only those phases
are physical\footnote{In the transition probability formulas for
neutrino oscillations ($\nu_\alpha \to \nu_\beta$)
$$
P(\nu_\alpha \to \nu_\beta) \equiv P_{\alpha\beta} \equiv
P_{\alpha\beta}(L) = \sum_{a=1}^3 \sum_{b=1}^3 
J_{\alpha\beta}^{ab} e^{i \frac{\Delta m_{ab}^2}{2E} L},
$$
where $\alpha,\beta = e,\mu,\tau$; $L$ is the neutrino baseline
length, $E$ is the neutrino energy, $J_{\alpha\beta}^{ab} \equiv
U^\ast_{\alpha a} U_{\beta a} U_{\alpha b} U^\ast_{\beta b}$ are the
amplitude parameters, and $\Delta m_{ab}^2 \equiv m_a^2 - m_b^2$ are the
neutrino mass squared differences, we observe that the matrix elements of the
leptonic mixing matrix $U = (U_{\alpha a})$, where $\alpha =
e,\mu,\tau$ and $a = 1,2,3$, only appear in the amplitude parameters
$J^{ab}_{\alpha\beta}$. It is obvious from the definitions of the
amplitude parameters 
$$
J_{\alpha\beta}^{ab} \equiv U^\ast_{\alpha a} U_{\beta a} U_{\alpha b}
U^\ast_{\beta b}
$$
to see that they are invariant under phase transformations of the following
kind
$$
U_{\alpha a} \to U'_{\alpha a} = e^{i \varphi_\alpha} U_{\alpha a}
e^{-i \varphi_a},
$$
where $\varphi_\alpha, \varphi_a \in \mathbb{R}$ are arbitrary
parameters, \ie, ${J'}_{\alpha\beta}^{ab} = J_{\alpha\beta}^{ab}$.
Thus, the transition probability formulas may only depend on phases in
the matrix $U$, which cannot be absorbed by the above phase
transformations (see \eq~(\ref{eq:U=PUP+}) for the corresponding
matrix form). The number of such phases is equal to $1$ (in the $3
\times 3$ case), \ie, the $\mathcal{CP}$ violation phase $\delta$
\cite{Bilenky:1980cx,Doi:1981yb}.},
which cannot be eliminated by the transformation
\begin{equation}
U \to U' = \Phi_\ell U \Phi_\nu^\dagger,
\label{eq:U=PUP+}
\end{equation}
where the matrices $\Phi_\ell \equiv {\rm diag \,}({\rm e}^{{\rm i}
  \varphi_e},{\rm e}^{{\rm i} \varphi_\mu},{\rm e}^{{\rm i} \varphi_\tau})$ and
  $\Phi_\nu \equiv {\rm diag \,}({\rm e}^{{\rm i} \varphi_1},{\rm e}^{{\rm i}
  \varphi_2},{\rm e}^{{\rm i} \varphi_3})$ can always be put on the
forms
\begin{eqnarray}
\Phi_\ell &=& {\rm e}^{{\rm i} \varphi_\ell} \Phi_{\ell'}, \\
\Phi_\nu &=& {\rm e}^{{\rm i} \varphi_\nu} \Phi_{\nu'}
\end{eqnarray}
such that $\det \Phi_{\ell'} = 1$ and $\det \Phi_{\nu'} = 1$, where
  $\Phi_{\ell'} \equiv {\rm diag \,}({\rm e}^{{\rm i}
  \varphi_{e'}},{\rm e}^{{\rm i}
  \varphi_{\mu'}},{\rm e}^{{\rm i} \varphi_{\tau'}})$ and $\Phi_{\nu'}
  \equiv {\rm diag \,}({\rm e}^{{\rm i} \varphi_{1'}},{\rm e}^{{\rm i} 
  \varphi_{2'}},{\rm e}^{{\rm i} \varphi_{3'}})$.
This means that $\varphi_\alpha = \varphi_\ell + \varphi_{\alpha'}$
  ($\alpha = e,\mu,\tau$) and $\varphi_a = \varphi_\nu + \varphi_{a'}$
  ($a = 1,2,3$), where $\varphi_{e'} + \varphi_{\mu'} +
  \varphi_{\tau'} = 0$ and $\varphi_{1'} + \varphi_{2'} + \varphi_{3'}
  = 0$.
Thus, we have
\begin{equation}
U \to U' = {\rm e}^{{\rm i} (\varphi_\ell - \varphi_\nu)} \Phi_{\ell'} U
\Phi_{\nu'} = {\rm e}^{{\rm i} \varphi} \Phi_{\ell'} U \Phi_{\nu'},
\end{equation}
where $\varphi \equiv \varphi_\ell - \varphi_\nu$, and on matrix form,
we find that
\begin{equation}
U' = {\rm e}^{{\rm i} \varphi} \left( \begin{matrix} U_{11} {\rm e}^{{\rm i}
    (\varphi_{e'} - \varphi_{1'})} & U_{12} {\rm e}^{{\rm i}
    (\varphi_{e'} - \varphi_{2'})} & U_{13} {\rm e}^{{\rm i}
    (\varphi_{e'} - \varphi_{3'})} \\ U_{21} {\rm e}^{{\rm i}
    (\varphi_{\mu'} - \varphi_{1'})} & U_{22} {\rm e}^{{\rm i}
    (\varphi_{\mu'} - \varphi_{2'})} & U_{23} {\rm e}^{{\rm i}
    (\varphi_{\mu'} - \varphi_{3'})} \\ U_{31} {\rm e}^{{\rm i}
    (\varphi_{\tau'} - \varphi_{1'})} & U_{32} {\rm e}^{{\rm i}
    (\varphi_{\tau'} - \varphi_{2'})} & U_{33} {\rm e}^{{\rm i}
    (\varphi_{\tau'} - \varphi_{3'})} \end{matrix} \right).
\label{eq:U'}
\end{equation}
Assuming now that the matrix $U'$ in \eq~(\ref{eq:U'}) is on the ``standard''
parameterization form as displayed in \eq~(\ref{eq:Ustand}), we identify
from the 1-1, 1-2, 1-3, 2-3, and 3-3 entries that
\begin{eqnarray}
C_2 C_3 &=& {\rm e}^{{\rm i} \varphi} U_{11} {\rm e}^{{\rm i} (\varphi_{e'} -
  \varphi_{1'})} = |U_{11}| {\rm e}^{{\rm i} (\arg U_{11} + \varphi +
  \varphi_{e'} - \varphi_{1'})}, \label{eq:e1}\\
S_3 C_2 &=& {\rm e}^{{\rm i} \varphi} U_{12} {\rm e}^{{\rm i} (\varphi_{e'} -
  \varphi_{2'})} = |U_{12}| {\rm e}^{{\rm i} (\arg U_{12} + \varphi +
  \varphi_{e'} - \varphi_{2'})}, \label{eq:e2}\\
S_2 {\rm e}^{- {\rm i} \delta} &=& {\rm e}^{{\rm i} \varphi} U_{13}
  {\rm e}^{{\rm i}
  (\varphi_{e'} - \varphi_{3'})} = |U_{13}| {\rm e}^{{\rm i} (\arg
  U_{13} + \varphi +   \varphi_{e'} - \varphi_{3'})}, \label{eq:e3}\\ 
S_1 C_2 &=& {\rm e}^{{\rm i} \varphi} U_{23} {\rm e}^{{\rm i}
  (\varphi_{\mu'} - \varphi_{3'})} = |U_{23}| {\rm e}^{{\rm i}
  (\arg U_{23} + \varphi + \varphi_{\mu'} - \varphi_{3'})}, \label{eq:m3}\\
C_1 C_2 &=& {\rm e}^{{\rm i} \varphi} U_{33} {\rm e}^{{\rm i}
  (\varphi_{\tau'} - \varphi_{3'})} = |U_{33}| {\rm e}^{{\rm i}
  (\arg U_{33} + \varphi + \varphi_{\tau'} - \varphi_{3'})}. \label{eq:t3}
\end{eqnarray}
Taking the imaginary parts of \eqs~(\ref{eq:e1}), (\ref{eq:e2}),
(\ref{eq:m3}), and (\ref{eq:t3}), we arrive at
\begin{eqnarray}
0 &=& \arg U_{11} + \varphi + \varphi_{e'} - \varphi_{1'}, \\
0 &=& \arg U_{12} + \varphi + \varphi_{e'} - \varphi_{2'}, \\
0 &=& \arg U_{23} + \varphi + \varphi_{\mu'} - \varphi_{3'}, \\
0 &=& \arg U_{33} + \varphi + \varphi_{\tau'} - \varphi_{3'},
\end{eqnarray}
which together with the two conditions
\begin{eqnarray}
\varphi_{e'} + \varphi_{\mu'} + \varphi_{\tau'} &=& 0, \\
\varphi_{1'} + \varphi_{2'} + \varphi_{3'} &=& 0
\end{eqnarray}
make up a system of equations. Note that the overall phase $\varphi$ can be
absorbed into the $\arg U_{ab}$'s, \ie, $\arg U_{ab} + \varphi \to
\arg U_{ab}$. The system of equations includes six
equations and six unknown quantities, \ie, $\varphi_{\alpha'}$
($\alpha = e,\mu,\tau$) and $\varphi_{a'}$ ($a = 1,2,3$), and thus, it
has a unique solution. This solution is
\begin{eqnarray}
\varphi_{e'} &=& \frac{1}{3} \left( - 2 \arg U_{11} - 2 \arg U_{12} -
\arg U_{23} - \arg U_{33} \right),\\
\varphi_{\mu'} &=& \frac{1}{3} \left( \arg U_{11} + \arg U_{12} -
\arg U_{23} + 2 \arg U_{33} \right),\\
\varphi_{\tau'} &=& - \varphi_{e'} - \varphi_{\mu'},\\
\varphi_{1'} &=& \frac{1}{3} \left( \arg U_{11} - 2 \arg U_{12} -
\arg U_{23} - \arg U_{33} \right),\\
\varphi_{2'} &=& \frac{1}{3} \left( - 2 \arg U_{11} + \arg U_{12} -
\arg U_{23} - \arg U_{33} \right),\\
\varphi_{3'} &=& - \varphi_{1'} - \varphi_{2'}.
\end{eqnarray}
Inserting the solution into \eqs~(\ref{eq:e1}) - (\ref{eq:t3}), we
obtain
\begin{eqnarray}
C_2 C_3 &=& |U_{11}|, \label{eq:e1'}\\
S_3 C_2 &=& |U_{12}|, \label{eq:e2'}\\
S_2 {\rm e}^{- {\rm i} \delta} &=& |U_{13}| {\rm e}^{{\rm i} (\arg
  U_{13} - \arg U_{11} - \arg U_{12} - \arg U_{23} - \arg U_{33})},
  \label{eq:e3'}\\ 
S_1 C_2 &=& |U_{23}|, \label{eq:m3'}\\
C_1 C_2 &=& |U_{33}|. \label{eq:t3'}
\end{eqnarray}
Thus, it follows from \eq~(\ref{eq:e3'}) by taking the real and
imaginary parts, respectively, that
\begin{eqnarray}
S_2 &=& |U_{13}| \quad \Rightarrow \quad \theta_{13} \equiv \theta_2 =
\arcsin S_2 = \arcsin |U_{13}|, \\
- \delta &=& - \arg U_{11} - \arg U_{12} + \arg U_{13} - \arg
U_{23} - \arg U_{33}.
\end{eqnarray}
Dividing \eq~(\ref{eq:e2'}) by \eq~(\ref{eq:e1'}), we find that
\begin{equation}
\frac{S_3}{C_3} = \left| \frac{U_{12}}{U_{11}} \right| \quad
\Rightarrow \quad \theta_{12} \equiv \theta_{3} = \arctan
\frac{S_3}{C_3} = \arctan \left| \frac{U_{12}}{U_{11}} \right|,
\end{equation}
and similarly, dividing \eq~(\ref{eq:m3'}) by \eq~(\ref{eq:t3'}), we
find that
\begin{equation}
\frac{S_1}{C_1} = \left| \frac{U_{23}}{U_{33}} \right| \quad
\Rightarrow \quad \theta_{23} \equiv \theta_{1} = \arctan
\frac{S_1}{C_1} = \arctan \left| \frac{U_{23}}{U_{33}} \right|.
\end{equation}
In summary, the parameters of the ``standard'' parameterization form
is given as
\begin{eqnarray}
\theta_{12} &=& \arctan \left| \frac{U_{12}}{U_{11}} \right|, \\
\theta_{13} &=& \arcsin |U_{13}|, \\
\theta_{23} &=& \arctan \left| \frac{U_{23}}{U_{33}} \right|, \\
\delta &=& \arg U_{11} + \arg U_{12} - \arg U_{13} + \arg
U_{23} + \arg U_{33}
\end{eqnarray}
in terms of the entries of any $3 \times 3$ unitary matrix $U =
(U_{ab})$. Note that these parameters are uniquely determined by
the entries in the first row and third column ($U_{11}$, $U_{12}$,
$U_{13}$, $U_{23}$, and $U_{33}$) only as well as the overall
phase $\varphi$. The other entries are completely restricted by and
follow directly from the unitarity conditions ($U^\dagger U = U
U^\dagger = 1_3$).

\end{appendix}


\begin{thebibliography}{10}

\bibitem{Cabibbo:1963yz}
N. Cabibbo,
\newblock Phys. Rev. Lett. 10 (1963) 531.

\bibitem{Kobayashi:1973fv}
M. Kobayashi and T. Maskawa,
\newblock Prog. Theor. Phys. 49 (1973) 652.

\bibitem{Dermisek:1999vy}
R. Dermisek and S. Raby,
\newblock Phys. Rev. D 62 (2000) 015007, {\tt hep-ph/9911275}.

\bibitem{King:2001uz}
S.F. King and G.G. Ross,
\newblock Phys. Lett. B 520 (2001) 243, {\tt hep-ph/0108112}.

\bibitem{Fukuda:1998mi}
Super-Kamiokande Collaboration, Y. Fukuda et~al.,
\newblock Phys. Rev. Lett. 81 (1998) 1562, {\tt hep-ex/9807003}.

\bibitem{Fukuda:1998ah}
Super-Kamiokande Collaboration, Y. Fukuda et~al.,
\newblock Phys. Rev. Lett. 82 (1999) 2644, {\tt hep-ex/9812014}.

\bibitem{Fukuda:2000np}
Super-Kamiokande Collaboration, S. Fukuda et~al.,
\newblock Phys. Rev. Lett. 85 (2000) 3999, {\tt hep-ex/0009001}.

\bibitem{Toshito:2001dk}
Super-Kamiokande Collaboration, T. Toshito,
\newblock {\tt hep-ex/0105023}.

\bibitem{Shiozawa:2002}
M. Shiozawa,
\newblock talk given at the XXth International Conference on Neutrino Physics
  \& Astrophysics (Neutrino 2002), Munich, Germany, 2002.

\bibitem{Fukuda:2001nj}
Super-Kamiokande Collaboration, S. Fukuda et~al.,
\newblock Phys. Rev. Lett. 86 (2001) 5651, {\tt hep-ex/0103032}.

\bibitem{Smy:2001wf}
Super-Kamiokande Collaboration, M.B. Smy,
\newblock {\tt hep-ex/0106064}.

\bibitem{Fukuda:2002pe}
Super-Kamiokande Collaboration, S. Fukuda et~al.,
\newblock Phys. Lett. B 539 (2002) 179, {\tt hep-ex/0205075}.

\bibitem{Smy:2002}
M. Smy,
\newblock talk given at the XXth International Conference on Neutrino Physics
  \& Astrophysics (Neutrino 2002), Munich, Germany, 2002.

\bibitem{Ahmad:2001an}
SNO Collaboration, Q.R. Ahmad et~al.,
\newblock Phys. Rev. Lett. 87 (2001) 071301, {\tt nucl-ex/0106015}.

\bibitem{Ahmad:2002jz}
SNO Collaboration, Q.R. Ahmad et~al.,
\newblock Phys. Rev. Lett. 89 (2002) 011301, {\tt nucl-ex/0204008}.

\bibitem{Ahmad:2002ka}
SNO Collaboration, Q.R. Ahmad et~al.,
\newblock Phys. Rev. Lett. 89 (2002) 011302, {\tt nucl-ex/0204009}.

\bibitem{Hallin:2002}
A. Hallin,
\newblock talk given at the XXth International Conference on Neutrino Physics
  \& Astrophysics (Neutrino 2002), Munich, Germany, 2002.

\bibitem{mikh85}
S.P. Mikheyev and A.Y. Smirnov,
\newblock Yad. Fiz. 42 (1985) 1441,
\newblock [Sov. J. Nucl. Phys. 42 (1985) 913].

\bibitem{mikh86}
S.P. Mikheyev and A.Y. Smirnov,
\newblock Nuovo Cimento C 9 (1986) 17.

\bibitem{wolf78}
L. Wolfenstein,
\newblock Phys. Rev. D 17 (1978) 2369.

\bibitem{Bahcall:2001zu}
J.N. Bahcall, M.C. Gonzalez-Garcia and C. Pe{\~n}a-Garay,
\newblock {J. High Energy Phys.} 08 (2001) 014, {\tt hep-ph/0106258}.

\bibitem{Bahcall:2001cb}
J.N. Bahcall, M.C. Gonzalez-Garcia and C. Pe{\~n}a-Garay,
\newblock {J. High Energy Phys.} 04 (2002) 007, {\tt hep-ph/0111150}.

\bibitem{Bahcall:2002hv}
J.N. Bahcall, M.C. Gonzalez-Garcia and C. Pe{\~n}a-Garay,
\newblock {J. High Energy Phys.} 07 (2002) 054, {\tt hep-ph/0204314}.

\bibitem{Barger:2002iv}
V. Barger et~al.,
\newblock Phys. Lett. B 537 (2002) 179, {\tt hep-ph/0204253}.

\bibitem{Bandyopadhyay:2002xj}
A. Bandyopadhyay et~al.,
\newblock Phys. Lett. B 540 (2002) 14, {\tt hep-ph/0204286}.

\bibitem{Aliani:2002ma}
P. Aliani et~al.,
\newblock {\tt hep-ph/0205053}.

\bibitem{deHolanda:2002pp}
P.C. de~Holanda and A.Y. Smirnov,
\newblock {\tt hep-ph/0205241}.

\bibitem{groo00}
Particle Data Group, D.E. Groom et~al.,
\newblock Eur. Phys. J. C 15 (2000) 1,
\newblock {\tt http://pdg.lbl.gov/}.

\bibitem{Ibanez:1994ig}
L.E. Iba{\~n}ez and G.G. Ross,
\newblock Phys. Lett. B 332 (1994) 100, {\tt hep-ph/9403338}.

\bibitem{Binetruy:1995ru}
P. Bin{\'e}truy and P. Ramond,
\newblock Phys. Lett. B 350 (1995) 49, {\tt hep-ph/9412385}.

\bibitem{Shafi:2000su}
Q. Shafi and Z. Tavartkiladze,
\newblock Phys. Lett. B 482 (2000) 145, {\tt hep-ph/0002150}.

\bibitem{Mohapatra:1998ka}
R.N. Mohapatra and S. Nussinov,
\newblock Phys. Rev. D 60 (1999) 013002, {\tt hep-ph/9809415}.

\bibitem{Wetterich:1998vh}
C. Wetterich,
\newblock Phys. Lett. B 451 (1999) 397, {\tt hep-ph/9812426}.

\bibitem{Grimus:2001ex}
W. Grimus and L. Lavoura,
\newblock J. High Energy Phys. 07 (2001) 045, {\tt hep-ph/0105212}.

\bibitem{Mohapatra:1999zr}
R.N. Mohapatra, A. P{\'e}rez-Lorenzana and C.A. de~Sousa~Pires,
\newblock Phys. Lett. B 474 (2000) 355, {\tt hep-ph/9911395}.

\bibitem{Fritzsch:1996dj}
H. Fritzsch and Z.z. Xing,
\newblock Phys. Lett. B 372 (1996) 265, {\tt hep-ph/9509389}.

\bibitem{Fritzsch:1998xs}
H. Fritzsch and Z.z. Xing,
\newblock Phys. Lett. B 440 (1998) 313, {\tt hep-ph/9808272}.

\bibitem{Fritzsch:1999ee}
H. Fritzsch and Z.z. Xing,
\newblock Prog. Part. Nucl. Phys. 45 (2000) 1, {\tt hep-ph/9912358}.

\bibitem{Fukugita:1998vn}
M. Fukugita, M. Tanimoto and T. Yanagida,
\newblock Phys. Rev. D 57 (1998) 4429, {\tt hep-ph/9709388}.

\bibitem{Fukugita:1998kt}
M. Fukugita, M. Tanimoto and T. Yanagida,
\newblock Phys. Rev. D 59 (1999) 113016, {\tt hep-ph/9809554}.

\bibitem{Tanimoto:1998yz}
M. Tanimoto,
\newblock Phys. Rev. D 59 (1999) 017304, {\tt hep-ph/9807283}.

\bibitem{Kang:1998gs}
S.K. Kang and C.S. Kim,
\newblock Phys. Rev. D 59 (1999) 091302, {\tt hep-ph/9811379}.

\bibitem{Tanimoto:1999pj}
M. Tanimoto, T. Watari and T. Yanagida,
\newblock Phys. Lett. B 461 (1999) 345, {\tt hep-ph/9904338}.

\bibitem{Dorsner:2001sg}
I. Dorsner and S.M. Barr,
\newblock Nucl. Phys. B 617 (2001) 493, {\tt hep-ph/0108168}.

\bibitem{King:2002}
S. King,
\newblock talk given at the XXth International Conference on Neutrino Physics
  \& Astrophysics (Neutrino 2002), Munich, Germany, 2002.

\bibitem{He:2002rv}
H.J. He, D.A. Dicus and J.N. Ng,
\newblock Phys. Lett. B 536 (2002) 83, {\tt hep-ph/0203237}.

\bibitem{Lindner:2001kd}
M. Lindner and W. Winter,
\newblock {\tt hep-ph/0111263}.

\bibitem{King:2002nf}
S.F. King,
\newblock {\tt hep-ph/0204360}.

\bibitem{Ohlsson:2002na}
T. Ohlsson and G. Seidl,
\newblock Phys. Lett. B 537 (2002) 95, {\tt hep-ph/0203117}.

\bibitem{barg98}
V. Barger et~al.,
\newblock Phys. Lett.~B 437 (1998) 107,
\newblock {\tt hep-ph/9806387}.

\bibitem{wein79}
S. Weinberg,
\newblock Phys. Rev. Lett. 43 (1979) 1566.

\bibitem{wilc792}
F. Wilczek and A. Zee,
\newblock Phys. Rev. Lett. 43 (1979) 1571.

\bibitem{frog79}
C.D. Froggatt and H.B. Nielsen,
\newblock Nucl. Phys.~B 147 (1979) 277.

\bibitem{Babu:2001ex}
K.S. Babu and C.N. Leung,
\newblock Nucl. Phys. B 619 (2001) 667, {\tt hep-ph/0106054}.

\bibitem{Froggatt:2002tb}
C.D. Froggatt, H.B. Nielsen and Y. Takanishi,
\newblock Nucl. Phys. B 631 (2002) 285, {\tt hep-ph/0201152}.

\bibitem{Nielsen:2002cw}
H.B. Nielsen and Y. Takanishi,
\newblock {\tt hep-ph/0205180}.

\bibitem{gell79}
M. Gell-Mann, P. Ramond and R. Slansky,
\newblock {\it Complex Spinors and Unified Theories}, in {\it Supergravity},
  Proceedings of the Workshop on Supergravity, Stony Brook, New York, 1979,
  edited by P. van Nieuwenhuizen and D.Z. Freedman (North-Holland, Amsterdam,
  1979), p. 315.

\bibitem{yana79}
T. Yanagida,
\newblock in {\it Proceedings of the Workshop on the Unified Theory and Baryon
  Number in the Universe}, edited by O. Sawada and A. Sugamoto (KEK, Tsukuba,
  1979), p. 79.

\bibitem{Mohapatra:1980ia}
R.N. Mohapatra and G. Senjanovi{\'c},
\newblock Phys. Rev. Lett. 44 (1980) 912.

\bibitem{Leurer:1993wg}
M. Leurer, Y. Nir and N. Seiberg,
\newblock Nucl. Phys. B 398 (1993) 319, {\tt hep-ph/9212278}.

\bibitem{Perez-Lorenzana:2001pr}
A. P{\'e}rez-Lorenzana and C.A. de~Sousa~Pires,
\newblock Phys. Lett. B 522 (2001) 297, {\tt hep-ph/0108158}.

\bibitem{Green:1984sg}
M.B. Green and J.H. Schwarz,
\newblock Phys. Lett. B 149 (1984) 117.

\bibitem{Krauss:1989zc}
L.M. Krauss and F. Wilczek,
\newblock Phys. Rev. Lett. 62 (1989) 1221.

\bibitem{Ibanez:1991hv}
L.E. Iba{\~n}ez and G.G. Ross,
\newblock Phys. Lett. B 260 (1991) 291.

\bibitem{Witten:2001bf}
E. Witten,
\newblock {\tt hep-ph/0201018}.

\bibitem{Sher:1989mj}
M. Sher,
\newblock Phys. Rept. 179 (1989) 273.

\bibitem{Witten:1981kv}
E. Witten,
\newblock Phys. Lett. B 105 (1981) 267.

\bibitem{Leurer:1994gy}
M. Leurer, Y. Nir and N. Seiberg,
\newblock Nucl. Phys. B 420 (1994) 468, {\tt hep-ph/9310320}.

\bibitem{Ohlsson:1999xb}
T. Ohlsson and H. Snellman,
\newblock J. Math. Phys. 41 (2000) 2768, {\tt hep-ph/9910546},
\newblock 42 (2001) 2345(E).

\bibitem{Ohlsson:2001vp}
T. Ohlsson,
\newblock Phys. Scripta T93 (2001) 18.

\bibitem{Maki:1962mu}
Z. Maki, M. Nakagawa and S. Sakata,
\newblock Prog. Theor. Phys. 28 (1962) 870.

\bibitem{Apollonio:1998xe}
CHOOZ Collaboration, M. Apollonio et~al.,
\newblock Phys. Lett. B 420 (1998) 397, {\tt hep-ex/9711002}.

\bibitem{Apollonio:1999ae}
CHOOZ Collaboration, M. Apollonio et~al.,
\newblock Phys. Lett. B 466 (1999) 415, {\tt hep-ex/9907037}.

\bibitem{Bemporad:1999de}
CHOOZ Collaboration, C. Bemporad,
\newblock Nucl. Phys. B (Proc. Suppl.) 77 (1999) 159.

\bibitem{Bilenky:1987ty}
S.M. Bilenky and S.T. Petcov,
\newblock Rev. Mod. Phys. 59 (1987) 671,
\newblock 61 (1989) 169(E).

\bibitem{Bilenky:2001rz}
S.M. Bilenky, S. Pascoli and S.T. Petcov,
\newblock Phys. Rev. D 64 (2001) 053010, {\tt hep-ph/0102265}.

\bibitem{Lindner:2002}
M. Lindner, T. Ohlsson and G. Seidl,
\newblock Phys. Rev. D 65 (2002) 053014, {\tt hep-ph/0109264}.

\bibitem{Pascoli:2002xq}
S. Pascoli and S.T. Petcov,
\newblock {\tt hep-ph/0205022}.

\bibitem{Klapdor-Kleingrothaus:2000dg}
H.V. Klapdor-Kleingrothaus,
\newblock Nucl. Phys. B (Proc. Suppl.) 100 (2001) 309, {\tt hep-ph/0102276}.

\bibitem{Aalseth:2000ud}
IGEX Collaboration, C.E. Aalseth et~al.,
\newblock Phys. Atom. Nucl. 63 (2000) 1225,
\newblock [Yad. Fiz. 63 (2000) 1299].

\bibitem{Aalseth:2002rf}
IGEX Collaboration, C.E. Aalseth et~al.,
\newblock Phys. Rev. D 65 (2002) 092007, {\tt hep-ex/0202026}.

\bibitem{Hannestad:2002xv}
S. Hannestad,
\newblock {\tt astro-ph/0205223}.

\bibitem{Hannestad:2002}
S. Hannestad,
\newblock talk given at the XXth International Conference on Neutrino Physics
  \& Astrophysics (Neutrino 2002), Munich, Germany, 2002.

\bibitem{Wang:2001gy}
X. Wang, M. Tegmark and M. Zaldarriaga,
\newblock Phys. Rev. D 65 (2002) 123001, {\tt astro-ph/0105091}.

\bibitem{Elgaroy:2002bi}
{\O}. Elgar{\o}y et~al.,
\newblock Phys. Rev. Lett. 89 (2002) 061301, {\tt astro-ph/0204152}.

\bibitem{Hu:1998mj}
W. Hu, D.J. Eisenstein and M. Tegmark,
\newblock Phys. Rev. Lett. 80 (1998) 5255, {\tt astro-ph/9712057}.

\bibitem{Bilenky:1980cx}
S.M. Bilenky, J. Hosek and S.T. Petcov,
\newblock Phys. Lett. B 94 (1980) 495.

\bibitem{Doi:1981yb}
M. Doi et~al.,
\newblock Phys. Lett. B 102 (1981) 323.

\end{thebibliography}
\end{document}